\documentclass[11pt]{article}

\usepackage[english,francais]{babel}

\usepackage[latin1]{inputenc}
\usepackage{epsfig,graphicx}
\usepackage{latexsym}

\newfont{\ensmathquatorze}{msbm10 scaled 1400}
\newfont{\ensmathonze}{msbm10 scaled 1100}
\newfont{\ensmathdix}{msbm10}
\newfont{\ensmathneuf}{msbm10 scaled 833}
\newfont{\ensmathhuit}{msbm10 scaled 694}
\newfam\ensmathfam                        
\textfont\ensmathfam=\ensmathonze        
\scriptfont\ensmathfam=\ensmathdix       
\scriptscriptfont\ensmathfam=\ensmathhuit

\def\be{\begin{equation}}
\def\ee{\end{equation}}
\def\bea{\begin{eqnarray}}
\def\eea{\end{eqnarray}}

\newcommand\lij{l_{ij}}
\newcommand\uij{u_{ij}}
\newcommand\zij{z_{ij}}
\newcommand\xij{x_{ij}}
\newcommand\xim{x_{im}}

\newcommand\cim{\chi_{im}}
\newcommand\cji{\chi_{ji}}

\newcommand\lab{l_{\alpha\beta}}
\newcommand\labi{l_{\alpha\beta_i}}
\newcommand\ab{\alpha\beta}
\newcommand\ba{\beta\alpha}
\newcommand\abi{\alpha\beta_i}
\newcommand\abc{\alpha\beta_1}
\newcommand\abma{\alpha\beta_{m_\alpha}}
\newcommand\xab{x_{\alpha\beta}}
\newcommand\yab{y_{\alpha\beta}}
\newcommand\uab{u_{\alpha\beta}}
\newcommand\zab{z_{\alpha\beta}}
\newcommand\xabi{x_{\alpha\beta_i}}

\newcommand\cab{\chi_{\alpha\beta}}
\newcommand\cba{\chi_{\beta\alpha}}

\newcommand\lla{\left\langle}
\newcommand\rra{\right\rangle}

\def\d{{\rm d}}

\def\fp{\displaystyle }

\begin{document}

\selectlanguage{english}

\title{ Exit and Occupation times for Brownian Motion on Graphs with
 General Drift and Diffusion Constant }

\author{ Olivier B\'enichou$^1$  and Jean Desbois$^2$  }

\maketitle	

{\small
\noindent
$^1$ Laboratoire de Physique   Th\'eorique  de la Mati\`ere
Condens\'ee, Universit\'e Pierre et Marie Curie,
4, place Jussieu, 75005 Paris, France.

\noindent
$^2$ Laboratoire de Physique Th\'eorique et Mod\`eles Statistiques.
Universit\'e Paris-Sud, B\^at. 100, F-91405 Orsay Cedex, France.

}

\begin{abstract}  

\noindent
We consider a particle diffusing along the links of a
general graph possessing some absorbing vertices. The particle, with
a spatially-dependent diffusion constant $D(x)$ is subjected to a drift
$U(x)$ that is defined in every point of each link.

\noindent
We establish the boundary conditions to be used at the vertices and
we derive general expressions for the average time spent on a part of the
graph before absorption and, also, for the Laplace transform of the joint
law of the occupation times. Exit times distributions
and splitting probabilities are also studied and several examples are
discussed.

\end{abstract}

\section{Introduction}\label{intro}

\vskip.3cm 

\noindent
For many years, graphs have interested physicists as well as mathematicians.
For instance, equilibrium statistical physics widely uses model systems
defined on lattices, the most popular being certainly the Ising model
\cite{bax}. On another hand, in solid-state physics, tight-binding models
 (see, for instance, \cite{eco}) involve discretized
 versions of Schr\"odinger operators on graphs.
 For all those models, the thermodynamic properties
of the system heavily depend on geometrical characteristics of the lattice
 such as the connectivity and the dimensionality of the embedding space.
However, in general, they don't depend explicitly on the lengths of the
edges. Random walks on graphs, where a particle hops from one vertex to one
of its nearest-neighbours, have also been studied by considering discrete
Laplacian operators on graphs \cite{cvet}.

\vskip.1cm 

\noindent 
Such Laplacian operators can also be useful if they are defined on each link
of the graph. For example, in the context of organic molecules \cite{ru},
 they can describe free electrons on networks made of one-dimensional wires.
Many other applications can be found in the physical literature. Let us
simply cite the study of vibrational properties of fractal structures
 such as the Sierpinski gasket \cite{ram} or the investigation of quantum
 transport in mesoscopic physics \cite{avr, texier}. Weakly disordered
 systems can also be studied in that context \cite{mont}. It appears that
weak localization corrections in the presence of an eventual magnetic field
are related to a spectral determinant on the graph. This last quantity is
 actually of central importance and interesting by itself \cite{akk, jd}.
 In particular, it allows to recover a trace formula that was first derived
 by Roth \cite{roth}. Moreover, the spectral determinant, when computed
with generalized boundary conditions at the vertices, is useful to enumerate
constrained random walks on a general graph \cite{jd1}, a problem that has  
been addressed many times in the mathematical literature \cite{ihara}.  

\vskip.1cm 

\noindent
Brownian motion on graphs is also worthwhile to be investigated from, both,
the physical and mathematical viewpoints. For instance, the probability
distribution of the time spent on a link (the so-called occupation time) was
first studied by P Levy \cite{levy} who considered the time spent on an
infinite half-line by a one-dimensional brownian motion  stopped
at some time $t$. This work allowed Levy to discover in 1939 one of his
famous arc-sine laws \cite{levy1}. Since that time, this result has been 
generalized to a star-graph \cite{yor} and also to a quite general graph
\cite{jd2}. Local time distributions  have also been obtained in 
 \cite{comtet}.

\vskip.1cm 

\noindent
It has been pointed out since a long time that first-passage times and, more
generally, occupation times are of special interest in the context of
 reaction-diffusion processes \cite{vk, red}. Computations of such quantities
 in the presence of a constant external field have already been performed
for one-dimensional   systems   with absorbing points (see, for example, \cite{agmon}).
 This was done with the help of a linear Fokker-Planck equation 
 \cite{vk, risken}.  

\vskip.1cm 

\noindent
The purpose of the present work is to extend those results on a general
graph with some absorbing vertices. We will consider a brownian particle
diffusing with a spatially-dependent diffusion constant and subjected 
 to a drift that is defined in every point of each link. The paper is
 organized as follows. In section 2, we present the notations that will
  be used throughout the paper. We discuss the boundary conditions
  to be used at each vertex in section 3 and, also, in the
  Appendices.
More precisely, we analyse in details specific graphs in the
Appendices A and B. The obtained results allow to deal with a general graph
  in Appendix C. 
Section 4 is devoted to the computation
of the average time spent, before absorption, by a brownian particle on a part
 of the graph. In this section, we also calculate the Laplace transform of the
 joint law of the occupation times on each link. In the following
 section, we present additional results, especially concerning conditional
 and splitting probabilities. Various examples are discussed all along the
 different sections. Finally, a brief summary is given in section 6. 

\vskip.3cm

\section{Definitions and notations}\label{defi}

\vskip.3cm 

\noindent
Let us consider a general
graph $ \cal G $ made of $V$ vertices
   linked by $B$ bonds of finite lengths.
On each bond [$\alpha\beta$], of length
    $l_{\alpha\beta }$, we define the coordinate $x_{\alpha\beta } $ that
    runs from $0$ (vertex $\alpha$) to $l_{\alpha\beta } $
     (vertex $\beta$). (We have, of course,  
 $x_{\beta\alpha } = l_{\alpha\beta } - x_{\alpha\beta } $).

\vskip.1cm 

\noindent
Moreover, we suppose that, among all the vertices,
 $N$ of them are absorbing. (A particle gets trapped if it reaches
  such a vertex).

\vskip.3cm

\noindent
We will study the motion on $ \cal G $ of a brownian particle that starts
at $t=0$ from some  non-absorbing point $x$. The particle with a
spatially-dependent
 diffusion constant $D(x)$ is subjected to a drift $U(x)$ defined on
  the bonds of $ \cal G $. More precisely, $D(x)$ and $U(x)$ are
  differentiable functions of $x$ on each link. In particular,
 on each link  [$\alpha\beta$], the following limits
 $D_{( \alpha\beta )} \equiv    \lim_{\xab  \to 0} D(\xab )$, 
       $D'_{( \alpha\beta )} \equiv    \lim_{\xab  \to 0} \frac{\partial
   }{\partial \xab }  D(\xab )$, 
 $U_{( \alpha\beta )} \equiv    \lim_{\xab  \to 0} U(\xab )$, ...,
are well defined. Such notations will be used extensively throughout
the paper.

\vskip.3cm

\noindent
The continuity properties of $D(x)$ and $U(x)$ at each vertex will be discussed in the
following section.

\vskip.1cm

\noindent
We also specify the motion  of the particle when it reaches some vertex
$\alpha $. Let us call $ \beta_i  $    ($i=1,\ldots ,m_\alpha $)
the  nearest neighbours of $\alpha $. We assume that the particle
 will come out towards $\beta_i$ with some
arbitrary  probability  $p_{\alpha\beta_i } $ 
 ($\sum_i p_{\alpha\beta_i } =1 $  -- see \cite{yor} for a rigourous
 mathematical definition). Of course,
$p_{\lambda\mu } = 0 $ if $\lambda $ is an absorbing vertex or
if $[\lambda\mu ]$ is not a bond of $ \cal G $.

\vskip.3cm

\noindent
Let $P(yt|x0)$ be the probability density to find the particle at point $y$
at time $t$ ($P(y0|x0) = \delta (y-x)$). It satisfies on each link
  [$\alpha\beta$]  the  backward and forward Fokker-Planck equations:

\bea
\frac{ \partial P (y t|\xab 0) }{ \partial t } & = &
D (\xab ) \frac{ \partial^2 P  }{ \partial \xab^2 } -
\frac{ \partial U (\xab ) }{ \partial \xab }
\frac{ \partial P   }{ \partial \xab } \equiv  L^+( \xab )P
 ( y t|\xab 0)  \label{fp}   \\
\frac{ \partial P  (\yab t|x 0)   }{ \partial t } & = &
\frac{ \partial }{ \partial \yab }\left[
\frac{ \partial }{ \partial \yab }(D(\yab )P) + 
\frac{ \partial U (\yab ) }{ \partial \yab } P
\right] \equiv  L( \yab )P(\yab t|x 0)   \label{fpforw} 
\eea

\vskip.3cm

\noindent
$P_{( \alpha \beta   )}$ will mean  $ \lim_{\xab \to 0} \; P(y t|\xab
0)$   when we use  the  backward Fokker-Planck equation and  $ \lim_{\yab \to 0} \; P(\yab t|x 0)$
  when we use  the forward  equation.
The derivatives $P'_{( \alpha \beta   )}$ will be defined in a
similar way.

\vskip.3cm 

\section{Boundary conditions}\label{bcond}

\vskip.3cm

\noindent
Let us define the following two situations that can occur at some
non absorbing   vertex $\alpha$:

\vskip.5cm

\begin{center}  

(A) \  \ : \ \ \ $D(x)$ is continuous but $U(x)$ is not

\vskip.4cm

(B) \ \ : \ \ \ $U(x)$ is continuous but $D(x)$ is not

\end{center}

\vskip.5cm

\noindent
The main purpose of this section is to establish the boundary
conditions for $P$ and its derivatives that result from those
discontinuities.
We will  not consider the case when both $U(x)$ and $D(x)$ are
discontinuous at the same vertex because, in our opinion, it is ill-defined.

\vskip.3cm 

\subsection{Backward Fokker-Planck equation}\label{bcbw}

\vskip.3cm

\begin{figure}
\begin{center}
\includegraphics[scale=.4,angle=0]{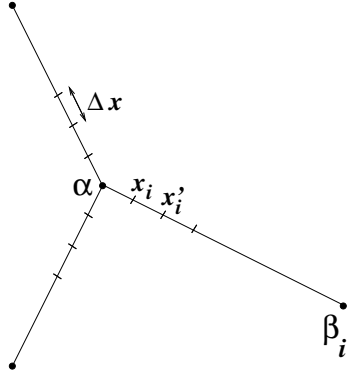}
\end{center}
\caption{The vertex $\alpha $ with its nearest-neighbours
  $\beta_i$, $i=1, ...,m_{\alpha }$; each link is discretized with
  steps of lengths $\Delta x$.   }\label{fig1}
\end{figure}

\noindent
Let us start by considering a graph ${\cal G}$ with  $U(x)$ and $D(x)$
constant (standard brownian motion). In Figure \ref{fig1}, we display a given
vertex $\alpha $ and its nearest-neighbours   $\beta_i$, $i=1,
...,m_{\alpha }$ on ${\cal G}$.

\vskip.1cm

\noindent
The transition probabilities from $\alpha
$, $p_\abi $, are not supposed to be all equal. In order to establish
the boundary conditions for the backward equation, we discretize all 
the links $[\abi ]$ (and also the time) with steps of length 
$\Delta x$ (resp. $\Delta t$). It is easy to realize that:

\bea
P(y, (N+1)\Delta t |  \alpha, 0   )         & = &  
 \sum_{i=1}^{m_{\alpha }  } \;  p_{\abi } \;  P ( y , N \Delta t |  x_i , 0   ) 
    \label{back1} \\
P ( y , (N+1) \Delta t |  x_i , 0   )        & = & 
\frac{1}{2}   P(y, N \Delta t |  \alpha, 0   ) +  \frac{1}{2}
  P ( y , N \Delta t |  x'_i , 0   ) 
     \label{back2}
\eea 

\vskip.3cm

\noindent 
Taking the limit  $\Delta x \to 0$, $\Delta t \propto (\Delta x)^2 $,
$N \to \infty$, $N \Delta t \equiv t$, we obtain, with (\ref{back2}):
\be\label{back3}
P_{(\abi )}= \frac{1}{2}   P(y, t |  \alpha, 0   ) +  \frac{1}{2} P_{(\abi )}
\ee
Thus  
\be\label{back30}
P_{(\abi )}=    P(y, t |  \alpha, 0   )  \qquad \forall i
\ee

\vskip.3cm

\noindent 
This shows that  $P$ is continuous in $\alpha$.

\vskip.3cm

\noindent    
Moreover, expanding (\ref{back1}) at order  $\Delta x$, we get:

\be\label{back4}
P(y, t |  \alpha, 0 ) = \sum_{i=1}^{m_{\alpha } }  p_{\abi} \;  P_{(\abi )} \;  +
\; \Delta x \left( \sum_{i=1}^{m_{\alpha } }  p_{\abi} \;  P'_{(\abi )}
\right) \; + \; O((\Delta x)^2)
\ee
With (\ref{back30}) and $\sum_{i=1}^{m_{\alpha } }  p_{\abi} =1$, we
show that:
\be\label{back5} 
\sum_{i=1}^{m_{\alpha } } \;  p_{\abi} \;  P'_{(\abi )} \; = \; 0  
\ee

\vskip.6cm

\noindent 
On the other hand, for $p_{\alpha\beta_1}=p_{\alpha\beta_2}=
...=1/m_{\alpha }$, the equation (\ref{fp})
on the link $[ \alpha\beta ]$ can be written in the
form:
\bea
\frac{\partial }{\partial t} \left(  \frac{1}{D(\xab )} e^{- \Phi
   (\xab ) } P(yt | \xab 0)   \right)
                 & = &   \frac{\partial }{\partial \xab} \left(   e^{-
\Phi (\xab ) } \frac{\partial }{\partial \xab }  P(yt | \xab 0) \right) \label{back6}  \\
\mbox{with} \qquad 
  \Phi (\xab )  & = & \int_{x_0}^{\xab }  \frac{\partial U(x')}{\partial
    x'} 
 \frac{\d x'}{ D(x')}      \label{back7} 
\eea
$x_0$ is some point on the graph.

\vskip.1cm

\noindent  
We are aware that $\Phi (\xab )$ could be multi-valued
because, in general, a graph is multiply-connected\footnote{
It can also occur that $\Phi(x)$ is not well defined, for instance, when $U(x)$ and 
 $D(x)$ are both discontinuous at the same point $x_1$. Indeed, 
$\fp \frac{1}{D(x')} \frac{\partial U(x') }  {\partial x'}    $ is then
proportional to $\delta (x'-x_1) / D(x_1) $; thus, this quantity is not defined if 
$D(x) $ is discontinuous at $x_1$.}. However, this is not the case if
we restrict ourselves to the vicinity of the given vertex
$\alpha$. Choosing $x_0$ located on a link starting from $\alpha$, the
integral in  (\ref{back7}) involves in a unique way, at most, two integrals along
links starting from  $\alpha$. It is well defined if $U(x)$ and $D(x)$
are not discontinuous at the same point. So, with this definition of
$\Phi(\xab )$, the  equation  (\ref{back6})   is well-suited to
study the boundary conditions at vertex  $\alpha$.

\vskip.6cm

\noindent 
Let us consider the case (A).

\vskip.6cm

\noindent
We assume, first, that the $p_{\abi}$'s are  all equal.

\vskip.2cm

\noindent 
Integrating  (\ref{back6}) in the vicinity of
$\alpha$, we get, with $D(x)$ continuous at $\alpha $:

\be\label{back8} 
\sum_{i=1}^{m_{\alpha } } \;   e^{- U_{(\abi   )} / D(\alpha )  }  \;  P'_{(\abi )} \; = \; 0  
\ee

\vskip.2cm

\noindent 
Moreover, $P$ must be continuous at $\alpha$ if we want  the quantity \ $\fp
e^{-\Phi} \frac{ \partial P} {\partial x   }   $ \ to be properly defined.

\vskip.2cm

\noindent
In the Appendix A, those boundary conditions are directly established,
for the Laplace Transform of $P$, on a simple graph.

\vskip.6cm

\noindent 
Still for the case (A), let us now  assume  that the $p_{\abi}$'s are not all
equal.

\vskip.4cm

\noindent 
In Appendix C, we establish the following boundary
condition that must hold for a general graph:
\be\label{back13} 
\sum_{i=1}^{m_{\alpha } } \; p_{\abi} \;  e^{-U_{(\alpha\beta_i )}
  /D(\alpha )} \; P'_{(\alpha\beta_i )}
 \; = \; 0
\ee

\vskip.2cm

\noindent 
We also show in this Appendix  that  $P$ is continuous in $\alpha$.

\vskip1.cm

\noindent
Turning to  the case (B), we can follow the same line as for (A)
and use (\ref{back6}) and also the Appendices A and B.

\vskip.3cm

\noindent
Remark that, when the $p_{\abi}$'s ($i=1,2,...,m_\alpha$) are all
equal, the integration of (\ref{back6}) on the vicinity  of  $\alpha$
will not produce an exponential factor as in (\ref{back8}). This is
because, this time, $U(x)$ is continuous in  $\alpha$.
This fact is confirmed by the direct computations performed in the
Appendices.
Finally, with the  $p_{\abi}$'s not all equal, we get: 
\be\label{back23} 
\sum_{i=1}^{m_{\alpha } } \; p_{\abi} \;  P'_{(\alpha\beta_i )}
 \; = \; 0
\ee

\vskip.3cm

\noindent
Moreover, $P$ is continuous for the cases (A) and (B).

\vskip.6cm

\noindent
In summary, for the backward  Fokker-Planck equation, the condition
on the derivatives can be
written:

\be\label{st9} 
\sum_{i=1}^{m_\alpha}  \; p'_{\abi } P'_{(\abi )}   \;  =   \;  0
\ee

\bea
\mbox{with} \qquad   p'_{\abi }    &=&  p_{\abi } e^{ -\frac{ U_{(
     \abi)} }  {D(\alpha )}  }    \hskip.5cm  \qquad \mbox{case \ (A)}   \label{st99}     \\
                                  &=&  p_{\abi }  \hskip.2cm   \qquad \qquad
                               \qquad    \mbox{case \ (B)}  \label{st999}
\eea

\vskip.3cm

\noindent
This notation will show especially useful in the following sections where the  backward  Fokker-Planck equation
is widely used.

\subsection{Forward Fokker-Planck equation}\label{bcfw}

\vskip.3cm

\noindent
Coming back to Figure \ref{fig1} (with  $x_i$, $x'_i$ and $\Delta x$ replaced
by   $y_i$, $y'_i$ and $\Delta y$), we consider the discretized version
of the forward equation with $D(x) $ and $U(x) $ constant  and, also, the 
 $p_{\abi } $'s not all equal. We have, on the link $[ \abi ]$  

\bea
P(y_i, (N+1)\Delta t |  x, 0   )         & = &  p_{\abi }  P ( \alpha
, N \Delta t |  x , 0   ) +   \frac{1}{2}   P(y'_i, N \Delta t |  x, 0   )
\label{forw1} \\
P ( \alpha  , (N+1) \Delta t |  x , 0   )        & = &
\frac{1}{2}  \sum_{i=1}^{m_{\alpha }  } \; P ( y_i , N \Delta t |  x , 0   ) 
\label{forw2}
\eea 

\vskip.3cm

\noindent 
With  the limit  $\Delta y \to 0$, $\Delta t \propto (\Delta y)^2 $,
$N \to \infty$, $N \Delta t \equiv t$,  (\ref{forw1}) leads to:
\be\label{forw3}
P_{(\abi )}=   p_{\abi }    P(\alpha   , t |  x, 0   ) +  \frac{1}{2} P_{(\abi )}
\ee

\vskip.3cm

\noindent 
Thus  
\be\label{st1}
\frac{P_{(\abc )}}{p_{\abc }} \; = \; \frac{P_{(\alpha\beta_2
   )}}{p_{\alpha\beta_2   }} \; = \quad ...   \quad = \;
\frac{P_{(\abma )}}{p_{\abma }}=  2  P( \alpha , t | x , 0   ) 
\ee

\vskip.3cm

\noindent 
We see that, in general, $P$ is not continuous in $\alpha$.

\vskip.3cm

\noindent    
Moreover, expanding (\ref{forw2}) at order  $\Delta y$, we get:

\be\label{forw4}
P(\alpha, t |  x, 0 ) =   \frac{1}{2}    \sum_{i=1}^{m_{\alpha } } 
\left( \;  P_{(\abi )} \;  +  \; \Delta y  P'_{(\abi )}  \right)
\; + \; O((\Delta y)^2)
\ee
With (\ref{st1}), we can write:
\be\label{forw5} 
\sum_{i=1}^{m_{\alpha } } \;  P'_{(\abi )} \; = \; 0  
\ee

\vskip.1cm

\noindent
So, the current conservation doesn't involve the $p_{\abi}$'s.

\vskip.3cm

\noindent
Now, for $p_{\alpha\beta_1}=p_{\alpha\beta_2}=
...=1/m_{\alpha }$, the equation (\ref{fpforw})
on the link $[ \alpha\beta ]$ can be written:
\bea
\frac{\partial }{\partial t}  P(\yab t | x 0)     & = &
     \frac{\partial }{\partial \yab} \left(   e^{- \Phi (\yab ) } 
\frac{\partial }{\partial \yab } \left( D(\yab ) e^{ \Phi (\yab ) } 
P(\yab t | x 0) \right) \right) \label{forw6}  \\
\mbox{with} \qquad 
  \Phi (\yab )  & = & \int_{y_0}^{\yab }  \frac{\partial U(y')}{\partial
    y'} 
 \frac{\d y'}{ D(y')}      \label{forw7} 
\eea
$y_0$ is some point on the graph; the discussion of section
\ref{bcbw} for the definition of  $\Phi $ is, of course,  still relevant.

\vskip.3cm

\noindent
Following the same lines as in section \ref{bcbw}, we get the  current conservation at each non absorbing
vertex $\alpha$:

\be\label{st0}
\sum_{i=1}^{m_\alpha} \left(
(DP)'+U'P
\right)_{(\abi )}= \; 0 
\ee
if $\alpha $ is not the starting point. Otherwise, the right-hand side
of (\ref{st0}) should be replaced by \ $-\delta (t)$.

\vskip.1cm

\noindent
(\ref{st0}) doesn't depend on the continuity properties of
$D(x)$  and $U(x)$.

\vskip.1cm

\noindent
Now, let us consider the case (A) and call $D(\alpha )$ the diffusion
constant  at $\alpha $.

\vskip.1cm

\noindent
When $ p_{\abc }= ...= p_{\abma }    $, following \cite{risken}, we
can show that

\be\label{st2} 
e^{\frac{U_{( \abc)}}  {D(\alpha )}     } P_{(\abc )} \; = \; 
e^{\frac{U_{( \alpha\beta_2 )}}  {D(\alpha )}     } P_{(\alpha\beta_2
 )}  \; = \quad ... \quad =  \;
e^{\frac{U_{( \abma)}}  {D(\alpha )}     } P_{(\abma )}
\ee

\vskip.1cm

\noindent
When the $ p_{\abi } $'s are not all equal, we use the same approach
as the one of  Appendix  C  and get:

\be\label{st3}
\mbox{(A)} \quad  \mbox{:} \qquad 
e^{\frac{ U_{( \abc)} }  {D(\alpha )}     } \frac{ P_{(\abc )} } {p_{\abc
 }}  \; =  \;
e^{\frac{U_{( \alpha\beta_2 )}}  {D(\alpha )}     }  \frac{
 P_{(\alpha\beta_2    )} } {p_{\alpha\beta_2 }} \;  =  \quad
... \quad =  \;
e^{\frac{U_{( \abma)}}  {D(\alpha )}     }  \frac{ P_{(\abma )}
}{p_{\abma }}
\ee

\vskip.1cm

\noindent
Similarly, for the case (B), we obtain:

\be\label{st4}
\mbox{(B)} \quad : \qquad 
 D_{(\abc )}  \frac{P_{(\abc )}}{p_{\abc }}  \; = \;
D_{(\alpha\beta_2 )}       \frac{P_{(\alpha\beta_2
    )}}{p_{\alpha\beta_2   }}  \; = \quad ...  \quad  =  \;
D_{(\abma )}       \frac{P_{(\abma )}}{p_{\abma }}
\ee

\vskip.1cm

\noindent
When $ p_{\abc }= ...= p_{\abma }    $,
it is worth to notice that the conditions (\ref{st3}) and  (\ref{st4})
 are simply obtained if we want the quantity 
$\fp \frac{\partial }{\partial \yab } \left( D(\yab ) e^{ \Phi (\yab ) } 
P(\yab t | x 0) \right)$, that appears  in (\ref{forw6}), to be properly defined.

\vskip.3cm

\noindent
In the following section, we will check the consistency of those boundary conditions (for the
Backward and the Forward equations) on several examples.

\vskip.3cm 

\section{Occupation times}\label{ot}

\subsection{Mean residence time}\label{mrt}

\vskip.3cm

\noindent
Let us first study the average time,  $\lla \tau (x)\rra $, spent
on a part $ \cal D $ of $ \cal G $ by  the particle before absorption.
 We have \cite{agmon}:
\be\label{ag}
\lla \tau (x)\rra = \; \int_0^\infty \; \d t \;
\int_{\cal D } \;  \d y \;  P(yt|x0)
\ee
($\lla \tau (x)\rra $ is the sum of 
the infinitesimal residence times  weighted by the probability of presence
in ${\cal D }$ at time $t$).

\vskip.1cm

\noindent
With (\ref{fp}), we get for $\lla \tau (x) \rra $ the equation:
\be\label{tau} 
L^+ ( \xab ) \lla \tau  ( \xab )   \rra =  - {\bf 1}_{\cal D }(\xab)
\ee  
$ {\bf 1}_{\cal D }(x) $ is the characteristic function of the domain
$ \cal D $: $ {\bf 1}_{\cal D }(x) \; = \; 1  $ if $x \in {\cal D }$, \ 
   $= \; 0  $ otherwise.

\vskip.1cm

\noindent 
The solution writes:
\be\label{sol} 
\lla \tau (\xab) \rra  = 
\int_0^{\xab }\d \uab \; \phi (\uab ) +
A_{\ab} \int_0^{\xab} \d \uab \; I (\uab ) + B_{\ab}
\ee  
with
\bea
\phi (\xab ) & = & \; - \;  I (\xab ) \left(
 \int_0^{\xab} \frac{ \d \zab }{  D( \zab )    }
\; \frac{ {\bf 1}_{\cal D }(\zab) }{ I (\zab ) }
\right) \label{1} \\
 I (\xab )   & = & \exp \left( \int_0^{\xab} 
  \frac{ \d \zab }{  D( \zab )    }
   \frac{\partial  U  }{\partial\zab   }   
    \right)  \label{2} 
\eea
The constants $A_{\ab}$ and $B_{\ab}$ are determined by imposing the
boundary  conditions at each vertex. Continuity implies:
\bea  
\lim_{{\xabi}\to 0} \lla \tau (\xabi) \rra & \equiv &
\lla \tau_\alpha \rra = B_{\abi} \label{cont1} \\
\mbox{and} \qquad
\lim_{{\xabi}\to \labi } \lla \tau (\xabi) \rra & \equiv &
\lla \tau_{\beta_i} \rra =
 K_{(\abi )} + A_{\abi}  J_{(\abi )}    +  B_{\abi}  \label{cont2}
\eea   
where
\bea 
 K_{(\ab )}  &  =   &  \int_0^{\lab} \d \uab \; \phi (\uab )
 \label{Kab}      \\
 J_{(\ab )}  &  =   &  \int_0^{\lab} \d \uab \; I (\uab )  \label{Jab} 
\eea
In general,  $K_{(\ab )} \ne  K_{(\ba )}$,  $J_{(\ab )} \ne J_{(\ba )}$.

\vskip.3cm

\noindent
Of course, $\fp \lla \tau_\lambda \rra = 0 $ if
$ \lambda $ is absorbing. So, we will determine
$\fp \lla \tau_\alpha \rra $ only for  $ \alpha $ 
non-absorbing. In such a vertex, the current conservation leads to (\ref{st9}-\ref{st999}):
\be\label{cc1} 
\sum_i  \; p'_{\abi } \; A_{\abi} \; = \; 0 
\ee
The equations for the  $ \fp \lla \tau_\alpha \rra $'s can be written in a
matrix form and finally the average time spent on
$ {\cal D }$, before absorption, by a particle starting from $\alpha $ 
is given by:
\be\label{res1} 
\lla \tau_\alpha \rra = 
\frac{ \det M_1}{ \det M}
\ee
$M$ and $M_1$ are two $(V-N)\times (V-N)$ matrices 
 with the elements:
\bea
 M_{ii }    &=& \sum_m \frac{p'_{im}}{ J_{(im)} } \label{M} \\
M_{ij}    &=& \; - \; \frac{p'_{ij}}{  J_{(ij)}}          \label{MM}
\eea
(In (\ref{M},\ref{MM}), $i$ and $j$  run only over non-absorbing vertices
but $m$ labels all kinds of vertices).

\vskip.3cm

\noindent
$M_1=M$ except for the $\alpha^{\mbox{th}}$ column:
\be\label{M1}
 \left( M_1 \right)_{i\alpha } = \; - \; 
 \sum_m p'_{im} \frac{ K_{(im)}}{J_{(im)}}
\ee

\vskip.1cm

\noindent
We observe that, for a graph without any absorbing vertex,
 we have $\sum_j M_{ij} =0  $  $ \forall i  $. Thus,  
$\det M$ vanishes
and $ \fp \lla \tau_\alpha \rra $ becomes infinite as expected.

\vskip.3cm

\noindent
Remark also that, when  $ {\cal D} =    {\cal G}$,
$\tau $ becomes the survival time or, accordingly, the
 first-passage time in any absorbing vertex.

\vskip.1cm

\noindent
\subsubsection{Examples}\label{exam1}

\vskip.1cm

\noindent
EXAMPLE 1.

\vskip.3cm

\noindent
As a first example, let us consider 
the case $ {\cal D} =    {\cal G}$
for the graphs of Figure \ref{fig2}:

\begin{figure}
\begin{center}
\includegraphics[scale=.6,angle=0]{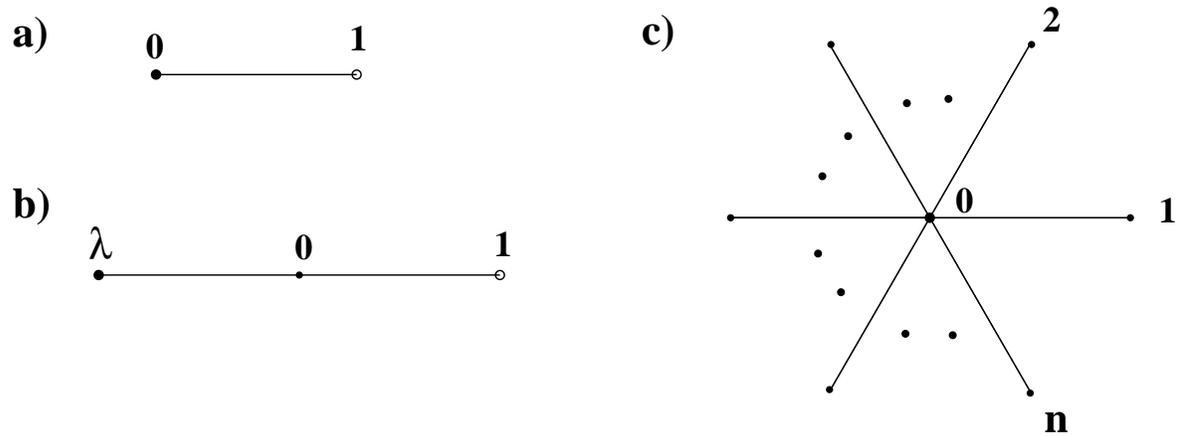}
\end{center}
\caption{  Three simple graphs that are used to compute survival times
(parts a) and b)) or covering time (c)).
In a) and b), the vertex $1$  is absorbing.
The brownian particle starts from $0$ (parts a) and c)) and from $\lambda$ 
 (part b)). For further explanations, see the  text. }\label{fig2}
\end{figure}

\vskip.1cm

\noindent
-- \ in a), the vertex $1$ is absorbing and the particle starts from $0$.
We get
\be\label{a}
 \lla \tau_0 \rra = \; - \; K_{(01)}
\ee

\begin{figure}
\begin{center}
\includegraphics[scale=.4,angle=0]{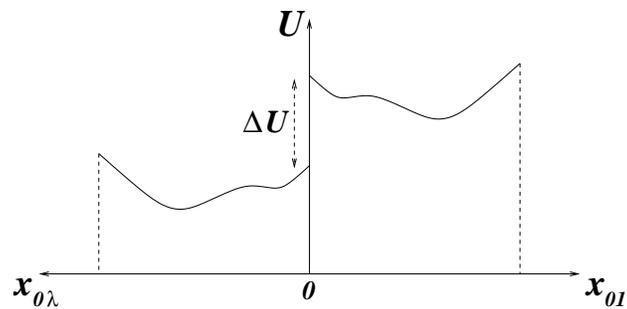}
\end{center}
\caption{   The potential is symmetric around $0$ except for a step of magnitude
 $\Delta U$.  }\label{fig3}
\end{figure}

\vskip.1cm

\noindent
-- \ in b), $1$ is still absorbing and the particle starts from
$\lambda $. Its probability transition in $0$ is $p_{01}=q$
($p_{0\lambda}=1-q$). Moreover, we assume that
 $l_{01}=l_{0\lambda}$,
$ U(x_{01})= U(x_{0\lambda})+ \Delta U$ (a step of magnitude $\Delta
U$   occurs in $0$, see Figure \ref{fig3}) and $ D(x_{01})= D(x_{0\lambda})$. The 
mean survival time is given by:

\be\label{b}
 \lla \tau_\lambda  (q) \rra  
     \; = \; - \; \left(
  \frac{K_{(01)}+K_{(10)}}{ q } \right)  \left(
q+(1-q)e^{\frac{\Delta U }    {D(0)}    }
\right)
\ee

\vskip.1cm

\noindent
--  \ c) represents a symmetric star with $n$
legs of length $l$ originating from
$0$ ($\fp p_{0i}= \frac{1}{n}  $ and  
 $ U(x_{0i})= U(x_{0j})$ (no step in $0$ this time) , $ D(x_{0i})= D(x_{0j})$
 $\forall i,j =1,\ldots ,n$).
Let us compute the average covering time $ \fp \lla \tau_c \rra$
(smallest time necessary to reach all the points of the star at 
least once).
 With the notations of (\ref{a}) and (\ref{b}), we get, with $\Delta
 U =0 $:
\bea  
 \lla \tau_c \rra & = & \lla \tau_0 \rra +
  \sum_{k=1}^{n-1} 
\lla \tau_\lambda (1-k/n)   \rra  = 
   \label{c1} \\
   & = & - K_{(01)} - (   K_{(01)}+K_{(10)}  ) 
   \; \;  n \;  \sum_{k=1}^{n-1} \frac{1}{k} 
    \label{c2} \\
  & \sim  &     - (   K_{(01)}+K_{(10)}  )     \; \; n \; \ln n
    \qquad \mbox{when} \ n\to\infty
  \label{c3} 
\eea
(Recall that for random walks on a star,
$\fp \lla \tau_c \rra \sim 2  \; n \; \ln n $
 \cite{Aldous})

\vskip.3cm

\noindent
EXAMPLE 2.

\vskip.3cm

\begin{figure}
\begin{center}
\includegraphics[scale=.5,angle=0]{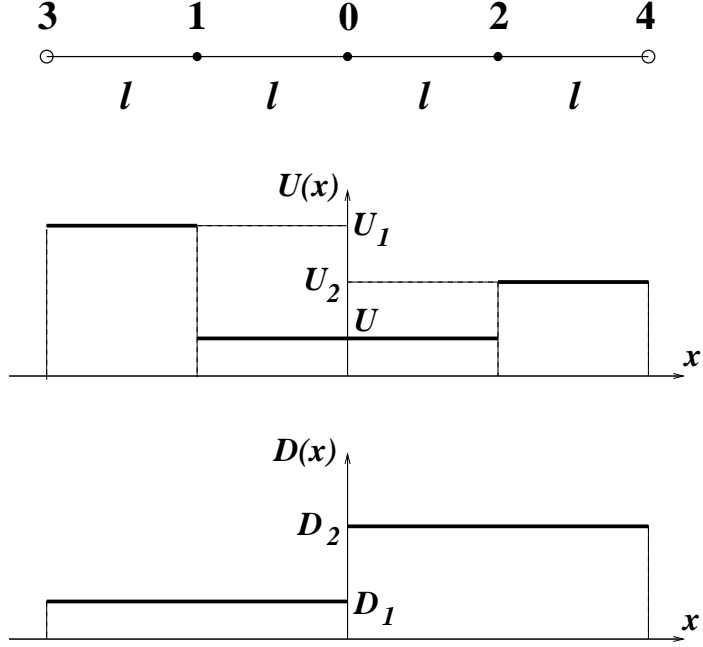}
\end{center}
\caption{  The  graph ${\cal G} $ consists in five vertices ($3$ and
 $4$: absorbing, $0$: starting point), $U(x)$ is discontinuous in $1$
 and $2$ and   $D(x)$ is discontinuous in $0$.       }\label{fig4}
\end{figure}

\noindent
Let us consider the graph of Figure \ref{fig4}. Among the five vertices, two are 
absorbing (3 and 4). The starting point is 0. $U(x) $ is
discontinuous in vertices 1 and 2,  $D(x) $ is
discontinuous in vertex 0. All the links have the same length $l$.
With the backward equation, we get the mean survival time
(\ref{res1}):
\be\label{examp2} 
\lla \tau_0 \rra    \; = \; l^2 \; 
\left( \frac{p_{01}}{D_1}  + \frac{p_{02}}{D_2}  \right)  \bigg/
\left( 
\frac{p_{01}p_{13}}{ p_{13} + p_{10} e^{(U_1 - U)/D_1}  }  +
\frac{p_{02}p_{24}}{ p_{24} + p_{20} e^{(U_2 - U)/D_2}  } 
 \right)
\ee

\vskip.3cm

\noindent
Let us now compute  $\lla \tau_0 \rra $ with the   forward equation.
Defining \ $ {\cal P} (y)  \equiv \int_0^\infty \d t \;  P(yt|00)  $, we
have
\bea
\mbox{on the link } [31] \; : \qquad {\cal P} (y_{31}) &=&  a_1 \;  
y_{31}+ a_2  \\
\mbox{on the link } [10] \; : \qquad {\cal P} (y_{01}) &=&  b_1 \;  
y_{01}+ b_2  \\
\mbox{on the link } [02] \; : \qquad {\cal P} (y_{02}) &=&  c_1 \;  
y_{02}+ c_2  \\
\mbox{on the link } [24] \; : \qquad {\cal P} (y_{42}) &=&  d_1 \;  
y_{42}+ d_2  
\eea 
with the boundary conditions
\bea
{\cal P}_{(31)}=0 \qquad {\cal P}_{(42)}=0 \qquad &&
\frac{D_1 {\cal    P}_{(01)} }{ p_{01} }  =   \frac{D_2 {\cal
    P}_{(02)}}  { p_{02} } \\
\frac{e^{U_1/D_1}   {\cal    P}_{(13)}   }{  p_{13} } = \frac{e^{U/D_1}   {\cal    P}_{(10)}  }{  p_{10} } 
  \qquad &&
\frac{e^{U_2/D_2}   {\cal    P}_{(24)}   }{  p_{24} } = \frac{e^{U/D_2}   {\cal    P}_{(20)}  }{  p_{20} }
\\
{\cal    P}'_{(13)} +  {\cal    P}'_{(10)} =0 \qquad &&
{\cal    P}'_{(24)} +   {\cal    P}'_{(20)} =0 \qquad 
 \qquad
D_2 {\cal    P}'_{(02)} +  D_1  {\cal    P}'_{(01)} =-1 \qquad 
\eea
Solving those conditions and computing $  \lla \tau_0 \rra =
\int_{ {\cal G} } {\cal P} (y) \d y = (b_2 +c_2)l$, we recover the result
equation (\ref{examp2}).

\vskip.3cm

\noindent
EXAMPLE 3.

\vskip.3cm

\begin{figure}
\begin{center}
\includegraphics[scale=.5,angle=0]{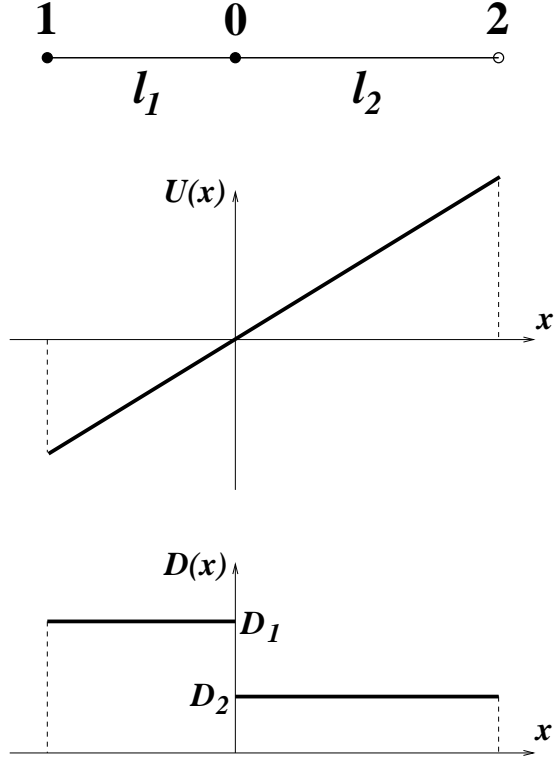}
\end{center}
\caption{   The  graph ${\cal G} $ consists in three vertices 
 ($2$: absorbing, $0$: starting point, pure reflection in 1); the drift $U(x)$ is linear 
   and   $D(x)$ is discontinuous in $0$.       }\label{fig5}
\end{figure}

\noindent
For the graph of Figure \ref{fig5}, the starting point is $0$, a pure reflection
occurs in $1$   ($p_{10}=1$) and the vertex $2$ is absorbing.
The potential is  
$U(x_{02})=  a \; x_{02} $, $U(x_{01})= - a \; x_{01} $. With the
notations of the Figure and using the backward result (\ref{res1}),
we get the mean residence time:
\bea
\mbox{ on the link} \quad [10] \qquad
\lla \tau_0 \rra & = & 
\frac{ p_{01}}{ p_{02}} \frac{D_2}{a^2} \left( e^{al_2/D_2} -1      \right)\left( e^{al_1/D_1} -1      \right)
\label{examp30}  \\
\mbox{ on the link} \quad [02] \qquad
\lla \tau_0 \rra & = & 
\frac{D_2}{a^2} \left( e^{al_2/D_2} -1 - \frac{a l_2}{D_2}
\right) \label{examp31}
\eea
Taking the limit $a\to 0$ (no drift), we obtain:
\bea
\mbox{ on the link} \quad [10] \qquad
\lla \tau_0 \rra & = &  \frac{p_{01}}{p_{02}} \frac{l_1l_2}{D_1} \label{examp32} \\
\mbox{ on the link} \quad [02] \qquad 
\lla \tau_0 \rra & = &  \frac{l_2^2}{2 D_2} \label{examp33} 
\eea

\vskip.1cm

\noindent
Turning to the forward equation, still with  $ {\cal P} (y)  \equiv
\int_0^\infty \d t \;  P(yt|00)  $, we have
\bea
\mbox{on the link } [10] \; : \qquad {\cal P} (y_{01}) &=& c_1 \;  
e^{ay_{01}/D_1}+c_2 \\
\mbox{on the link } [02] \; : \qquad {\cal P} (y_{02}) &=&  c_3 \;  
e^{- a  y_{02}/D_2}+ c_4
\eea 
with the conditions
\bea
{\cal P}_{(20)}=0 \qquad  \frac{D_1 {\cal    P}_{(01)} }{ p_{01} }  =   \frac{D_2 {\cal
    P}_{(02)}}  { p_{02} }
\qquad &&  D_1 {\cal    P}'_{(10)} + a \;  {\cal    P}_{(10)} =0  \\
D_2 {\cal    P}'_{(02)}  +  D_1 {\cal    P}'_{(01)} + a \; \left(
  {\cal    P}_{(02)} -  {\cal    P}_{(01)}      \right) = -1  
\eea
Computing $  \lla \tau_0 \rra =\int_{ {\cal D} } {\cal P}(y)  \;  \d y $ \
$ \; =
c_1\frac{D_1}{a}\left(   e^{al_1/D_1} -1   \right) +c_2l_1  \; $ for the
link $  [10]   $, \ 
$ \; =  c_3\frac{D_2}{a} \left( 1  -  e^{-al_2/D_2}  \right) +c_4l_2  \;  $ 
for the link $  [02]   $, \ 
  we recover the solution (\ref{examp30},\ref{examp31}).

\vskip.3cm 

\subsection{Occupation times distribution}\label{otd}

\vskip.3cm 

\noindent
Still with the same conditions (drift, spatially-dependent diffusion
constant, absorbing vertices,\ldots ),
we now study a Brownian motion stopped at $t$ and call
$T_{\ab}$ the time spent up to $t$ on the link [$\ab$].

\noindent
$T_{\ab}$ is, of course, a random variable depending on $t$
 and we will call ${\cal P}_t(\{ T_{\alpha\beta } \} )$
 the joint distribution of the  $T_{\alpha\beta }$'s.


\noindent
In the following, we will focus our attention on the quantity 
\bea
S(t|x0) &=&   \int  \prod_{[\alpha\beta ]} \d 
  T_{\alpha\beta} \;  
  {\cal P}_t(\{ T_{ \alpha\beta   } \} ) \;
  e^{- \sum_{[ij ]}\xi_{ij } 
 T_{ij } }  \label{defs1}  \\ 
 & \equiv &  \lla  e^{ -\sum_{[ij]}  \xi_{ij}T_{ij}}(x)  \rra 
   \label{defs2}
\eea 
where the $\xi_{ij}$'s are positive constants and
$\fp \lla  \ldots  (x) \rra $ stands for  averaging over all
the brownian trajectories  starting from $x$ and developing up to
 $t$ in the presence of the drift.

\vskip.1cm 

\noindent
In order to stick to $S(t|x0)$,
we slightly modify the backward equation (\ref{fp}) by adding,
 on each link [$\ab$], a loss term
proportional to $\xi_{\ab}$ ($\xi_{\ab}$ may be interpreted 
 as a reaction rate per unit length and time). Thus, we 
 consider the equation:  
\be\label{fpm} 
\frac{ \partial Q}{ \partial t } = 
 \left(   L^+(x_{\ab}) - \xi_{\ab} \right)  Q
\ee
where $Q(yt|x0)$ represents the probability of finding the particle
in $y$ at time $t$, each path from $x$ to $y$
being, this time, weighted by a factor \ 
$\fp e^{ -\sum_{[\ab]}  \xi_{\ab}T_{\ab}} $. (Notice that
 $Q(yt|x0)=\delta (y-x)$ if $x$ is absorbing).

\vskip.1cm 

\noindent  
It is now easy to realize that we have the relationship:
\be\label{s1}
S(t|x0)  =  \int_{\cal G} \d y \; Q(yt|x0) 
\ee

\vskip.1cm 

\noindent
In the following, we will be especially interested in 
the Laplace Transform of $S(t|x0)$:
\be\label{lt}
\widehat{S}(\gamma|x0) \equiv \widehat{S}(x) = 
 \int_0^\infty \d t \;  e^{-\gamma t} \; S(t|x0)
\ee

\vskip.1cm 

\noindent
On the bond [$\ab$], 
$\widehat{S}(\xab)$ satisfies the following equation
($\gamma_{\ab} \equiv \gamma + \xi_{\ab}$):
\be\label{lteq}
\left(   L^+(x_{\ab}) - \gamma_{\ab} \right)  \widehat{S} 
 = - 1
\ee

\vskip.1cm 

\noindent
Setting  $\fp \widehat{S}(\xab) = \frac{1}{\gamma_{\ab}} + \Phi (\xab) $
and performing the transformation
\be\label{tr} 
 \Phi (\xab) = \sqrt{I(\xab)} \; \chi (\xab)
\ee
($I$ is defined in (\ref{2})), we are left with the following equation
for $\chi$:
\be\label{eqchi} 
- D \frac{ \partial^2   \chi   }{ \partial \xab^2 } +
\left[
\frac{1}{4D}\left(
   \frac{ \partial U (\xab ) }{  \partial \xab  }
\right)^2
- D \frac{ \partial }{ \partial \xab}\left( 
\frac{1}{2D} \frac{ \partial U (\xab ) }{  \partial \xab  }
\right) + \gamma_{\ab } \right] \chi \;  = \; 0
\ee

\vskip.1cm 

\noindent
Let us call $\cab$ and $\cba$ two solutions of (\ref{eqchi})  such that
$\cab (\alpha)= \cba (\beta)=1$, $\cba (\alpha)= \cab (\beta)=0$.
 So,  $\fp \widehat{S} $ writes:
\bea
\widehat{S} = \frac{1}{\gamma_{\ab}}  & + & 
A_{\ab}
\exp \left( \int_0^{\xab} 
  \frac{ \d \zab }{ 2  D( \zab )    }\frac{ \partial U (\zab ) }{  \partial \zab  }
               \right) \cab (\xab) \nonumber \\
 &   +  &
A_{\ba}
\exp \left( \int_{\lab}^{\xab} 
  \frac{ \d \zab }{ 2  D( \zab )    }\frac{ \partial U (\zab ) }{  \partial \zab  }
    \right) 
\cba (\xab)    \label{sol2}
\eea
The constants $A_{ij}$ are determined by imposing 
the boundary conditions.
Continuity of $\widehat{S}$ at each vertex $\alpha$ implies:
\be\label{c10} 
\lim_{{\xabi} \to 0} \widehat{S}(\xabi) \equiv \widehat{S}_\alpha
=  \frac{1}{\gamma_{\abi}} + A_{{\abi}} 
\ee
Moreover, if $\alpha$ is absorbing then
$T_{\lambda\mu}=0 \ \forall \ [\lambda\mu ]$
and, from (\ref{defs1}) and (\ref{lt}), we get: 
\be\label{c11}  
\widehat{S}_\alpha =  \frac{1}{\gamma } = 
 \frac{1}{\gamma_{\abi}} + A_{{\abi}} 
\ee 

\vskip.1cm 

\noindent
Following the same lines as for $\fp \lla \tau_\alpha \rra$,
it is easy to show that  the $\widehat{S}_\alpha$'s 
 ($\alpha$ non-absorbing) satisfy a matrix equation with the solution:

\be\label{salph}  
\widehat{S}_\alpha =\frac{\det R_1}{\det R}
\ee
$R$ and $R_1$ are again $(V-N)\times (V-N)$ matrices with elements
\bea
  R_{ii}  &=& \sum_m p'_{{im}} C_{{im}}     \label{Raa} \\
  R_{ij}  &=&  p'_{ij} W_{ij}    \label{Raaa}
\eea

\noindent
the quantities $C_{im}$ and $W_{ij}$ being given by:
\bea
C_{im}  &=&  \lim_{{\xim} \to 0} \left(
\frac{\partial \cim}{ \partial \xim }+ 
\frac{  
 \frac{\partial U } {  \partial x_{im}  } }
{2 D(\xim )}
\right)
\label{Cab}  \\
W_{ij} &=& \exp\left(  \int_{\lij}^0 \frac{
\frac{ \partial U }{  \partial x_{ij}  }
}{2 D(\xij)}
 \d \xij \right) \lim_{{\xij} \to 0}\frac{\partial \cji}{ \partial \xij }
 \label{Wab}
\eea
$R_1=R$ except for the $\alpha^{\mbox{th}}$ column:
\be\label{R1ba}
\left( R_1  \right)_{i\alpha} = \sum_j 
\frac{p'_{ij}}{\gamma_{ij}} \left( C_{ij} +
W_{ij} \right) - \frac{1}{\gamma } 
\; \sum_{k \; \mbox{abs.}} \;  p'_{ik}
W_{ik}
\ee
(The last summation is performed only over absorbing vertices).

\vskip.1cm 

\noindent
Setting $\xi_{ij}\equiv 0$, we establish that  
$\fp \det R_1 = \frac{1}{\gamma} \det R$, showing that the probability
 distribution ${\cal P}_t(\{ T_{\alpha\beta } \} )$
 is properly normalized (see (\ref{defs1}) and (\ref{lt})).

\vskip.3cm 

\noindent
Now, let us call $T_{\ab}'$ the time spent on [$\ab$] before absorption.
Thus, for a particle starting from $\alpha$, we can write  
\be\label{abs}
\lla  e^{ -\sum_{[ij]} \xi_{ij}
T_{ij}'}(\alpha )  \rra =
\lim_{t\to \infty}
\lla  e^{ -\sum_{[ij]}  \xi_{ij}
T_{ij}}(\alpha )  \rra
\ee
and from (\ref{defs2},\ref{lt}) we deduce
\be\label{abs1}
\lla  e^{ -\sum_{[ij]}  \xi_{ij}
T_{ij}'}(\alpha )  \rra =
\lim_{\gamma\to 0} \left(  \gamma  \widehat{S}_\alpha \right)
\ee

\vskip.2cm

\noindent
Finally, we get:
\be\label{abs2}
\lla  e^{ -\sum_{[ij]}  \xi_{ij}
T_{ij}'}(\alpha )  \rra =
\frac{\det R_1^{(0)} }{ \det R^{(0)} }
\ee
with $\fp  R^{(0)} = \lim_{\gamma\to 0} R  $ and
$   R_1^{(0)}  = R^{(0)}   $ except for the $\alpha^{\mbox{th}}$
column that is given by 
\be\label{abs3} 
\left( R_1^{(0)}  \right)_{i\alpha} = - \lim_{\gamma\to 0} 
\; \sum_{k \; \mbox{abs.}}  
\; p'_{ik} W_{ik}
\ee
(\ref{abs3}) holds because
$$ \fp
\lim_{\gamma \to 0 } \left( \frac{\gamma }{\gamma_{ij}}p'_{ij} \left(
C_{ij}+W_{ij} \right) \right) = 0
$$
This is obvious when $\xi_{ij}\ne 0$. But when $\xi_{ij}= 0$, this is
still true because, in that case,  
 $\fp \lim_{\gamma \to 0}  \left( C_{ij}+W_{ij} \right) =0     $.

\vskip.3cm

\noindent
Remark that (\ref{abs3}) implies that 
\ $\fp  \lla  e^{ -\sum_{[ij]}  \xi_{ij}
T_{ij}'}(\alpha )  \rra \; = \; 0  $
\ if $\cal G$ has no absorbing vertex (in that case, $T'_{ij} = \infty$).

\vskip.1cm

\noindent
Setting $\xi_{ij} \equiv \xi$ and
$\fp \sum_{[ij]} T'_{ij}  \equiv \tau $ (survival time),
 (\ref{abs2}) gives the expression of the Laplace Transform of the
  probability distribution of $\tau$.

\vskip.1cm

\noindent
\subsubsection{Example}\label{exam2}

\vskip.3cm

\noindent
To illustrate this work, let us go back to the example 
of Figure \ref{fig2} and study the survival time
(parts a) and b) -- with $\Delta U =0$) and covering time (part c)) distributions.

\vskip.1cm

\noindent
With a), we get 
\be\label{A}
\lla  e^{ - \xi \tau_0}  \rra
= \; - \; \frac{W_{01}}{C_{01}}
\ee
($W_{\ab}$, $C_{\ab}$,\ldots are computed with $\gamma_{\ab}=\xi$).  

\vskip.1cm

\noindent
b) leads to 
\be\label{B}
\lla  e^{ - \xi \tau_\lambda    } (q) \rra 
= \; - \; \frac{q}{Z-1+q}
\ee
with $\fp Z= \frac{C_{01} C_{10}}{W_{01} W_{10}}$

\vskip.1cm

\noindent
For the covering time $\tau_c$ in c) we get (with the notations of
 (\ref{A}) and   (\ref{B})):
\bea
\lla  e^{ - \xi \tau_c}  \rra
& = & \; - \;  \lla  e^{ - \xi \tau_0}  \rra \;
\prod_{k=1}^{n-1} \lla  e^{ - \xi \tau_\lambda    } (1-k/n)
\rra 
  \label{C1}    \\
& = &  \; - \; \frac{W_{01}}{C_{01}} \;
\frac{(n-1)!}{n^{n-1}} \; \prod_{k=1}^{n-1} \; \frac{1}{Z- \frac{k}{n}}
\label{C2}
\eea
Computation of the first moments shows that when \ $n\to\infty$:
\bea
\lla  \tau_c    \rra & \propto & n \ln n 
 \qquad \mbox{in agreement with } \ (\ref{c3}) \label{c33} \\
 \lla  \tau_c^2    \rra   -  
\lla  \tau_c    \rra^2   & \propto & n^2    
\label{c34}
\eea
Then the  probability distribution of  the scaling variable $  \tau_c
/    \lla  \tau_c \rra
$ becomes more and more
peaked at its mean value when $n\to\infty$:
\be\label{peak} 
P\left( \frac{\tau_c}{\lla \tau_c \rra} = X \right)
\; \to_{n\to\infty} \;   \delta(X-1)
\ee

\vskip.3cm

\section{First passage times}

\vskip.3cm

\noindent
This section is essentially devoted to a detailed study of exit
(or survival) times. First, we will compute exit times distributions   
and, after, in the last parts, we will focus
on quantities concerning   exit  through a given absorbing vertex.

\vskip.3cm

\subsection{Exit times distribution}\label{etdistri}

\vskip.3cm

\noindent
The density of exit time $ {\cal P}(t|x0) $
(without specifying the absorbing vertex)
is defined through the loss of probability:

\bea
{\cal P}(t|x0) \;  \d t &=&  \int_{{\cal G}} \; \d y \; P(yt|x0)
- \int_{{\cal G}} \; \d y \; P(y,t+\d t|x0)   \\
\mbox{Thus:} \quad 
 {\cal P}(t|x0)  &=& -\frac{\partial   }{\partial t}
\int_{{\cal G}} \; \d y \; P(yt|x0)  \; \equiv \; 
\sum_{\mu \; \mbox{abs.}} {\cal P}_\mu (t|x0) \label{smu}
\eea
$\fp {\cal P}_\mu (t|x0)$ is the density of exit time by the absorbing vertex
$\mu$. 

\vskip.3cm

\noindent
With the forward Fokker-Planck equation, (\ref{smu}) leads to:

\bea
{\cal P}(t|x0) &=& - \int_{{\cal G}} \; \d y \;  L(y) P(yt|x0) =
 \sum_{\mu \; \mbox{abs.},\; i} J_{\mu \beta_i}  \\
J_{\mu \beta_i} &=& \lim_{y_{\mu \beta_i} \to 0} 
\left(   \frac{ \partial   } {\partial   y_{\mu \beta_i} } (D 
    P ) +
\frac{ \partial U  } {\partial   y_{\mu \beta_i} }  P 
\right)       \\
&=& \lim_{y_{\mu \beta_i} \to 0}  \left(    
D  \frac{ \partial  P  } {\partial   y_{\mu \beta_i} }
\right) 
\eea
because  $P_{(\mu\beta_i)}=0$ when $\mu$ is absorbing.
(The $\beta_i$'s are the nearest-neighbours of $\mu$ on ${\cal G}$.)

\vskip.1cm

\noindent
We deduce that $\fp {\cal P}_\mu (t|x0)$ is the   probability current
at vertex $\mu$: $\fp {\cal P}_\mu (t|x0) =  \sum_i  J_{\mu \beta_i}   $.

\vskip.3cm

\noindent
Let us compute the Laplace transform,
$\fp \widehat{{\cal P}} (\gamma |x0)$,   of  ${\cal P}(t|x0)$.
With (\ref{smu}), we get:
\bea
\widehat{{\cal P}} (\gamma |x0)  &=& \int_0^\infty \d t e^{-\gamma t} \left(
-\frac{\partial   }{\partial t}
\int_{{\cal G}} \; \d y \; P(yt|x0) \right) \equiv 
\sum_{\mu \; \mbox{abs.}} \widehat{{\cal P}}_{\mu} (\gamma |x0) \\
   &=& 1 - \gamma \int_0^\infty \d t e^{-\gamma t}
 \int_{{\cal G}} \; \d y \; P(yt|x0)
\eea

The backward Fokker-Planck equation gives:
\bea
L^+ \widehat{{\cal P}} (\gamma |x0) &=& \gamma \widehat{{\cal P}} (\gamma |x0) \\
\mbox{and} \quad L^+ \widehat{{\cal P}_{\mu}} (\gamma |x0) &=& \gamma \widehat{{\cal P}}_{\mu} (\gamma |x0) 
\label{smu1}
\eea

\vskip.1cm

\noindent
Defining  $ \fp  \widehat{{\cal P}}_{\mu , \alpha} = \lim_{
x_{ \alpha    i} \to 0 }    \widehat{{\cal P}}_{\mu} (\gamma |x_{\alpha
i}  0)    $, we remark that 
\be\label{pmulam}
\fp  \mbox{ for an absorbing vertex } \lambda  : \qquad
\widehat{{\cal P}}_{ \mu , \lambda } \;  = \; \delta_{ \mu  \lambda } 
\ee
\bea
\mbox{This is because}  \qquad  {\cal P}_{ \mu }(t| \lambda   0) &=&
\delta ( t ) \qquad \mbox{  if } \quad
   \lambda  = \mu  \\
&=&   0   \qquad \mbox{     otherwise }
\eea

\vskip.3cm

\noindent
Following section \ref{otd}, we readily find, for a particle
 starting from $\alpha$ (non absorbing), the Laplace transform
of the exit (by the vertex $\mu$) time distribution:
\be\label{jres1} 
\widehat{{\cal P}}_{\mu, \alpha} = 
\frac{ \det R_2^{(\mu, \alpha)}}{ \det R}
\ee
where $R_2^{(\mu,\alpha)}=R$ except for the $\alpha^{\mbox{th}}$ column:
\be\label{R2}
 \left( R_2^{(\mu,\alpha)} \right)_{i\alpha } = \;  
 -p'_{i\mu} W_{i \mu}
\ee
$R$ and $W_{i \mu}$ are defined in (\ref{Raa}-\ref{Wab}) and computed,
here, with $\xi_{\alpha\beta } \equiv 0     $.  ($p'_{i\mu}$ is defined
in (\ref{st99}-\ref{st999})).

\vskip.3cm

\vskip.3cm

\subsection{Splitting probabilities}\label{split} 

\vskip.3cm

\noindent
For a particle starting at $x$, we define 
the splitting probability $\Pi_\mu (x)$
  as the probability of ever reaching the absorbing
vertex $\mu$ (rather than any other absorbing vertex). Such a quantity
has already been considered for one-dimensional systems
\cite{Gardiner}  (see also \cite{Olivier} for extensions to
higher-dimensional systems).

\vskip.3cm

\noindent
We have, obviously:

\be\label{pimu}
\Pi_{ \mu } (x)  = \int_0^\infty \d t \; {\cal P}_{ \mu }(t| x   0) =
\widehat{{\cal P}}_{\mu} ( \gamma =0 | x  0)  
\ee

\vskip.1cm

\noindent
Setting $ \gamma = 0  $ in (\ref{smu1}) we see that 
 $ \Pi_{ \mu } (x) $ satisfies,  on $[\ab ]$, the backward equation:
\be\label{spl}
  L^+ (\xab) \Pi_\mu (\xab) = 0 
\ee

\noindent
The probability, $\Pi_{\mu, \alpha }$, for a particle
starting from the vertex $\alpha$ to be absorbed by $\mu$, is defined as 
  $\fp \Pi_{\mu, \alpha} = \lim_{x_{\alpha i} \to 0 } 
 \Pi_\mu (x_{\alpha i} ) $.

\vskip.1cm

\noindent
With  (\ref{pimu})   and  (\ref{pmulam}), it is easy to realize that    

\be\label{pimulam}
\fp  \mbox{ for an absorbing vertex } \lambda  : \qquad
 \Pi_{ \mu , \lambda } \;  = \; \delta_{ \mu  \lambda } 
\ee

\vskip.2cm

\noindent
Following the same lines as previously (see section \ref{mrt}), we find that $\Pi_{\mu,\alpha}$
($\alpha$ non absorbing) is again written as  
the ratio of two determinants:

\be\label{splittingres1} 
\Pi_{\mu,\alpha} = 
\frac{ \det M_2^{(\mu, \alpha)}}{ \det M}
\ee
where 
$M_2^{(\mu, \alpha)}=M$ except for the $\alpha^{\mbox{th}}$ column:
\be\label{M2}
 \left( M_2^{(\mu, \alpha)} \right)_{i\alpha } = \;  
 \frac{p'_{i\mu}}{J_{(i\mu)}}
\ee
($M$ is defined in (\ref{M},\ref{MM}) and $J_{(i\mu)}$ in (\ref{Jab})).

\vskip.2cm

\noindent
With simple manipulations on determinants, we check the normalization
condition  \ $\fp \sum_{i \; \mbox{abs.}} \Pi_{i, \alpha}=1$.

\vskip.5cm

\noindent
Let us, for one moment, comment the case when there is no drift 
($U(x)$ constant but $D(x)$ variable, eventually discontinuous at
some vertices). In that case $p'_{ij} = p_{ij}       $
and, also, $J_{(ij)} = l_{ij} $. We conclude that the splitting
probabilities don't depend on the varying diffusion constant when
there is no drift. This fact can be understood in the following
way. Let us consider a discretization of each link and a continuous
time. Modifying the diffusion constant amounts to change the waiting
time at each site of the discretized graph. But, this would not
change the trajectories if there is no drift. Only the time spent is
changed. Finally, the splitting probabilities remain unaffected\footnote{
  For a graph ${\cal G}$ without drift,   we could   expect,  with the same argument,
   that the average time spent on a part ${\cal D}$ of 
${\cal G}$ would not depend on the diffusion constant on the rest,
  ${\cal G} \backslash   {\cal D}  $, of  the graph. 
 This is exactly  what  can be checked with the formulae
 (\ref{res1}-\ref{M1}) of section  \ref{mrt} and, also, with   the
 formulae   (\ref{examp32},\ref{examp33}) of the example 3 of section  \ref{exam1}.}
  by a change of $D(x)$.

\vskip.1cm

\noindent
\subsubsection{Example}\label{exam3} 

\vskip.3cm

\noindent 
Let us consider a star-graph without drift. The root $0$ has $m_0$
neighbours, all absorbing. The $m_0$ links  have lengths $l_{0i}$, 
 $i=1,...,m_0$. With (\ref{splittingres1}), we obtain
\be\label{starsplit}
\Pi_{i,0} =
\left(
\frac{p_{0i}}{l_{0i}}
\right)
\bigg/
\left(
\sum_{m=1}^{m_0}
\frac{p_{0m}}{l_{0m}}
\right) 
\ee

\vskip.1cm

\noindent
\subsection{Conditional mean first passage times}\label{condit}

\vskip.3cm

We now turn to the study of the conditional mean first passage
time $\lla t_\mu(x)\rra $, which is defined as the mean exit time,
given  that exit is 
through the absorbing vertex $\mu$  (rather than any other absorbing
vertex).  We set   
$\fp \lla t_{\mu, \alpha}\rra = \lim_{x_{\alpha i} \to 0} 
 \lla t_\mu(x_{\alpha i})\rra 
$.

\vskip.2cm

\noindent
Actually, it is simpler to first compute 
the quantity
$\displaystyle \theta_\mu (x)\equiv\Pi_\mu (x)\lla t_\mu (x)\rra$.  

\vskip.2cm

\noindent
Indeed, we have:
\bea
\theta_\mu (x) &=&  \int_0^\infty \d t \; t \; {\cal P}_{ \mu } (t| x   0)   \\
  L^+  \theta_\mu (x)     &=&  \int_0^\infty \d t \; t \;
 \frac{\partial }{\partial t } {\cal P}_{
    \mu } (t| x   0) = -
  \int_0^\infty \d t  \; {\cal P}_{ \mu } (t| x   0)  \\   
\mbox{So, we get: } \quad   L^+ \theta_\mu (x)  &=& - \; \Pi_\mu (x)
\eea

\vskip.1cm

\noindent
Moreover, for any absorbing vertex $\lambda$, we get: 
$\fp \theta_{\mu , \lambda }= \Pi_{\mu , \lambda }
\lla t_{\mu,   \lambda }\rra =
    \delta_{\mu , \lambda } \lla t_{\mu,   \mu }\rra =        0$
because      $ \fp \lla t_{\mu,   \mu }\rra =        0 $.

\vskip.2cm

\noindent
Thus, comparing this equation with eq.(\ref{tau}), we find
  for a particle starting from the vertex $\alpha$ 
\be\label{tcond} 
\Pi_{\mu, \alpha}\lla  t_{\mu,\alpha}\rra = 
\frac{ \det M_3^{(\mu,\alpha)}}{ \det M}
\ee
where $M_3^{(\mu,\alpha)}=M$   except for the $\alpha^{\mbox{th}}$ column:
\be\label{M3}
 \left( M_3^{(\mu, \alpha)} \right)_{i\alpha } \; = \; 
 \sum_m p'_{im} \frac{ \widetilde{K}^{(\mu)}_{(im)}}
 {J_{(im)}}
\ee
with
\bea 
\widetilde{ K}^{(\mu)}_{(ij)}  &  =   &  \int_0^{\lij} \d \uij \;
I (\uij )  \left(
  \int_0^{\uij} \d \zij \; \frac{ \Pi_\mu (\zij ) }{ D(\zij) I (\zij ) }
\right) \
\eea
In this last equation, $ \Pi_\mu (\zij ) $ has to be computed by equation
\be
\Pi_\mu (\zij ) =  \Pi_{\mu , i} +\frac{ \Pi_{\mu , j} - 
\Pi_{\mu , i}}{J_{(ij)}}\;\int_0^{\zij} \d \uij \; I(\uij)
\ee
with   $ \Pi_{\mu , i}$ given by eq.(\ref{splittingres1}).

\vskip.1cm

\noindent
\subsubsection{Example}\label{exam4} 

\vskip.3cm

\noindent 
With the same star-graph as in section \ref{exam3} (no drift)
and a diffusion constant equal to  $D_{0i}$ on each link
 $[0i]$, we get (\ref{starsplit},\ref{tcond})

\be\label{startcond}
\lla t_{\mu, 0}   \rra \; = \; 
\frac{1}{6} \frac{l_{0\mu }^2}    {D_{0\mu }} \; + \; 
 \frac{1}{3} 
   \left( \sum_{m=1}^{m_0}
   p_{0m}       \frac{ l_{0m} } { D_{0m} }
\right)
\bigg/
\left( \sum_{m=1}^{m_0}
    \frac{ p_{0m} } { l_{0m} }
\right)
\ee

\section{Summary}\label{summary}

\vskip.3cm

\noindent
We have used the Fokker-Planck Equation (mainly the backward one)
to compute some quantities relevant for the study of a Brownian
particle moving on a graph. This particle, moving with a varying
diffusion constant, is also subjected to a drift.

\vskip.1cm

\noindent
Being aware that this point is somewhat controversial (see, for
instance, \cite{Torq}), we have
discussed in great details the boundary conditions in vertices where 
either $U(x)$ or $D(x)$ is discontinuous. Two Appendices support the
study of those boundary conditions.

\vskip.1cm

\noindent
Those preliminaries allowed us to deal with various quantities such as
the mean residence time, the occupation times probabilities,
the splitting probabilities and the conditional mean first passage
time. In each case, we have established general formulae that can be
used for all kinds of graphs.
\vskip.1cm

\noindent
Finally, we have checked, on several examples, the consistency of the
boundary conditions  we have put forward.

\vskip2.5cm

\begin{appendix}

\section{Backward Equation: boundary conditions when $p_{01} =p_{02}$ }\label{AA}

\noindent
For a direct computation of the boundary conditions, it is simpler to
consider the Laplace Transform:
\be\label{A01}
 \widehat{S} (\gamma | x0) \; = \; \int_0^\infty \d t \; e^{-\gamma
   t} \int_{{\cal G}} \d y P(yt |x0) 
\ee
that satisfies the backward equation ($ \widehat{S} (\gamma | x0)
\equiv  \widehat{S}(x)$):
\be\label{A02}
  ( L^+ - \gamma ) \widehat{S} (x) \; = \; -1
\ee

\vskip.5cm

\noindent
First, we want to study the case (A)

\vskip.3cm

\begin{figure}
\begin{center}
\includegraphics[scale=.5,angle=0]{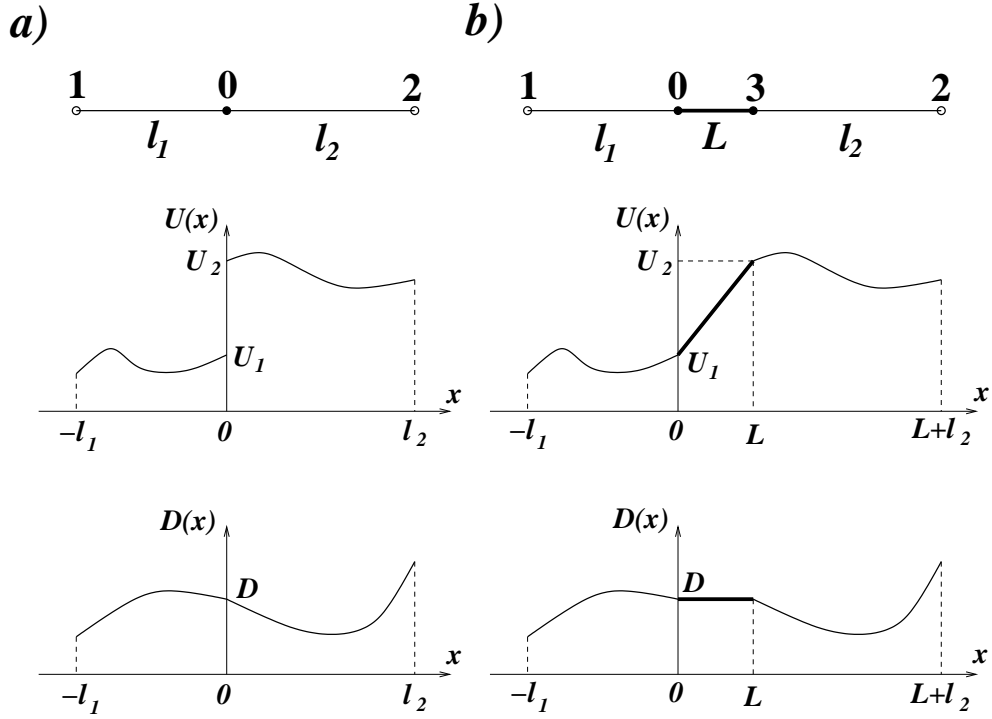}
\end{center}
\caption{ In part a) a simple graph with the absorbing vertices $1$
and $2$; the potential is discontinuous in $0$.
 In part b), we add the link $[03]$ (length $L$) and, on this link, a potential
 that interpolates linearly between $U_1$ and $U_2$; $D(x)$ is constant on
  $[03]$.  }\label{fig6}
\end{figure}

\noindent
For the simple graph of Figure \ref{fig6} a), 
$D(x)$ is continuous and $U(x)$ is 
discontinuous at the  vertex $0$. 
In particular  $U_{(01)} \equiv U_1 \ne  U_{(02)} \equiv U_2; D(0)
\equiv D    $. Moreover, vertices  $1$  and  $2$  are absorbing and,
in $0$, the transition probabilities are equal: $p_{01}=p_{02}=1/2$.

\noindent
Now, let us add to this graph, the link $[03]$
 of length $L$ (see Figure \ref{fig6} b))  in such a way that
nothing is changed for the rest (for example, the potential on the
link  $[02]$ of the original graph   is the same as the one on the
link $[32]$ of the modified graph, ...). On  $[03]$, we define
$U(x_{03})=  \left( \frac{U_2-U_1}{L} \right) x_{03}     +  U_1  $
and  $D(x_{03}) \equiv D $. When $ L \to 0 $, we recover  Figure \ref{fig6} a).

\vskip.5cm   

\noindent
Working with this modified graph, the solution of (\ref{A02}) writes:

\bea
\mbox{on the link } \; [10]  \qquad  \widehat{S}(x_{10}) & = & 1/\gamma + 
   a_1 \phi_1(x_{10})  +  a_2 \phi_2(x_{10})
  \label{A03}   \\    
\mbox{on the link } \; [03]   \qquad \widehat{S}(x_{03}) & = &  1/\gamma + 
 c_1 e^{r_+x_{03}} + c_2 e^{r_-x_{03}} 
   \label{A04}   \\
\mbox{on the link } \; [32]  \qquad  \widehat{S}(x_{32}) & = &   1/\gamma +   
b_1 \psi_1(x_{32})  +  b_2 \psi_2(x_{32})
\label{A05}   
\eea 
with
\bea
r_\pm  & = & \frac{A \pm \sqrt{ A^2 +4\gamma D    }}{ 2D  }     \label{A06}\\
A    & = & \frac{ U_2 -U_1 }{ L }  \label{A07}
\eea
$\phi_{1,2}$ and $\psi_{1,2}$ are solutions of the equation  $(L^+ -
\gamma) \phi = 0$ that satisfy   ($\phi_{(\alpha\beta )}\equiv \lim_{
 \xab \to 0} \phi (\xab )    $):

\bea
\phi_{1 \; (10)}=0 \quad  \phi_{1 \; (01)}=1 && \phi_{2 \; (10)}=1
\quad  \phi_{2 \; (01)}=0    \label{A08}       \\
\psi_{1 \; (32)}=1 \quad  \psi_{1 \; (23)}=0 && \psi_{2 \; (32)}=0
\quad  \psi_{2 \; (23)}=1  \label{A09}
\eea 

\vskip.5cm   

\noindent
$U(x)$ and $D(x)$ are continuous everywhere on the graph. So, we write
standard boundary conditions at the vertices (continuity of
$\widehat{S}$ and its derivative at vertices $0$ and $3$;
$\widehat{S}=0$ at vertices $1$ and $2$). We obtain
($\phi'_{(\alpha\beta )}\equiv \lim_{
 \xab \to 0} \frac{\partial \phi (\xab )}{ \partial  \xab   }    $):
\bea
a_2= b_2=-\frac{1}{\gamma   } \qquad a_1=c_1+c_2 && b_1= c_1
e^{r_+L}+c_2 e^{r_-L}  \\
a_1 \phi'_{1 \; (01)} +  a_2 \phi'_{2 \; (01)} +  c_1r_+ +c_2r_- & =
&  0 \\
b_1 \psi'_{1 \; (32)} +  b_2 \psi'_{2 \; (32)} -  c_1r_+ e^{r_+L}  -
c_2r_- e^{r_-L} & = &  0 
\eea

\vskip.5cm   

\noindent
After some algebra, we get the results:
\bea
\frac{ \widehat{S}_{(32)}     }{ \widehat{S}_{(01)}      }  & = &
\frac{ \frac{1}{\gamma} + b_1    }  { \frac{1}{\gamma} + a_1  }
\to_{L\to 0} \; 1    \label{A11}     \\
\frac{ \widehat{S}'_{(32)}     }{ \widehat{S}'_{(01)}      }  & = &
\frac{  b_1 \psi'_{1 \; (32)} +  b_2 \psi'_{2 \; (32)}   }  { a_1 \phi'_{1 \; (01)} +  a_2 \phi'_{2 \; (01)}  }
\to_{L\to 0} \; - \;  e^{(U_2 -U_1)/D}  \label{A12}
\eea
When $L\to 0$, the vertex $3$ moves to $0$ and equations (\ref{A11})
and (\ref{A12}) give, for the original graph:
\bea
\widehat{S}_{(01)}   & = &    \widehat{S}_{(02)}   \label{A13}     \\
\widehat{S}'_{(01)} e^{  - U_{(01)}  /  D(0)  }  & + & 
   \widehat{S}'_{(02)} e^{  - U_{(02)}  /  D(0)  }        = 0  \label{A14}
\eea 
In particular, we observe that $\widehat{S}$ is continuous at $0$ and
that exponential factors appear in the condition involving the derivatives.

\vskip.5cm   

\noindent
Let us now turn to  the case (B), still with the same graph and  $p_{01} = p_{02}     =
1/2$. 

\vskip.3cm   

\begin{figure}
\begin{center}
\includegraphics[scale=.5,angle=0]{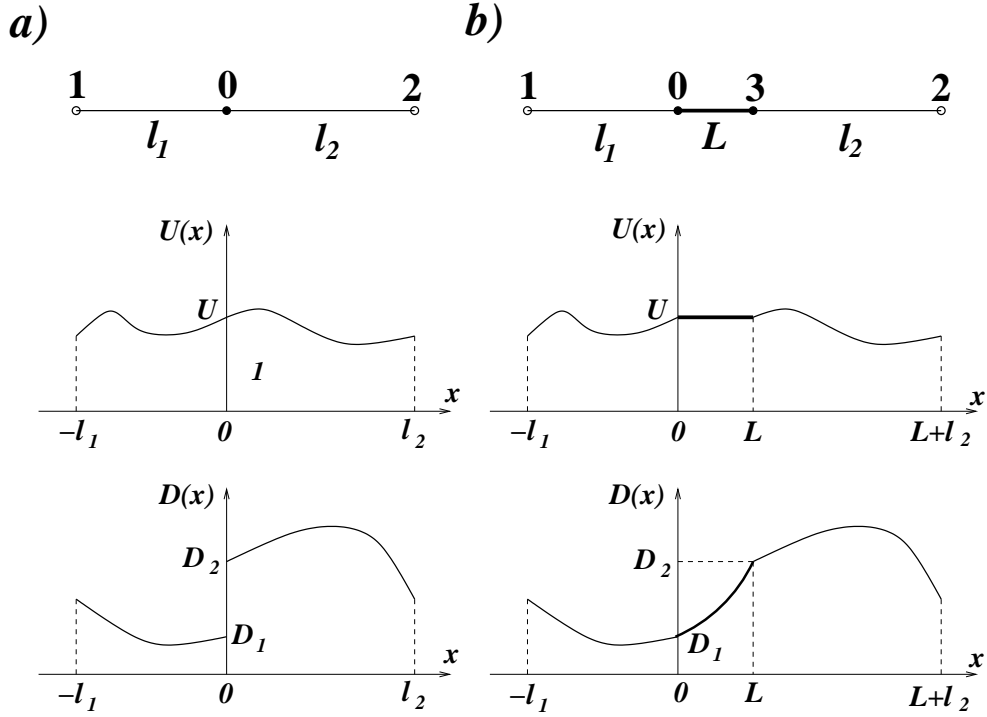}
\end{center}
\caption{ a) The same graph as in Figure \ref{fig6} a) but, this time, $D(x)$ is
 discontinuous in $0$; in part b), a parabolic interpolation is used
 to make  $D(x)$ continuous; $U(x)$ is  constant    on  the  link  $[03]$.  }\label{fig7} 
\end{figure}

\noindent
This time (Figure \ref{fig7} a)), $D(x)$ is discontinuous at $0$ and
$U(x)$ is continuous. We set: $D_{(01)}\equiv D_1 \ne D_{(02)}\equiv
D_2 $, $U(0) \equiv U $.

\vskip.5cm   

\noindent
As we already did for the case (A), we modify the
graph and obtain Figure \ref{fig7} b). Between vertices $0$ and $3$, we choose:
$U(x_{03})=U$ and $ \fp   D(x_{03}) = \left(\left( \frac{ \sqrt{D_2} - \sqrt{D_1}
  } {L}\right) x_{03} + \sqrt{D_1}    \right)^2   \; \equiv \; (a
x_{03} +b )^2  $. Thus, in the modified graph, $D(x)$ and $U(x)$ are
continuous everywhere on the graph.

\vskip.5cm   

\noindent
On  the links
$[10]$ and  $[32]$,  
the solution of equation (\ref{A02}) is still given by (\ref{A03}) and (\ref{A05}) 
 (with the conditions (\ref{A08})   and (\ref{A09})).
 But,  on   the link  $[03]$, (\ref{A04}) has to be
replaced by 

\bea
  \qquad \widehat{S}(x_{03}) &  = &  1/\gamma + 
 c_1  \; (a \; x_{03} + b)^{\lambda_+}    +  c_2  \; (a \; x_{03} +
 b)^{\lambda_-} \label{A20}  \\
 \mbox{with:} \qquad \lambda_\pm    &  = &    \frac{1}{2} \pm  \frac{1}{2} \sqrt{
   1+\frac{4\gamma }{a^2} } \label{A21}  \\
   a        &  = &   \frac{   \sqrt{D_2}  -   \sqrt{D_1}        }{ L
   } \label{A22}
\eea   
Standard boundary conditions imply:
\bea
a_2= b_2=-\frac{1}{\gamma   } \qquad 
a_1=c_1  D_1^{ \lambda_+/2 } + c_2 D_1^{ \lambda_-/2   } \qquad   
 b_1=c_1  D_2^{ \lambda_+/2 } + c_2 D_2^{ \lambda_-/2   }  &&  \\
a_1 \phi'_{1 \; (01)} +  a_2 \phi'_{2 \; (01)} + a\left(
 c_1   \lambda_+     D_1^{ (\lambda_+ - 1)/2 }  +
 c_2   \lambda_-     D_1^{ (\lambda_- - 1)/2 } \right) &  =  &  0 \\
b_1 \psi'_{1 \; (32)} +  b_2 \psi'_{2 \; (32)} - a\left(
 c_1   \lambda_+     D_2^{ (\lambda_+ - 1)/2 }  +
 c_2   \lambda_-     D_2^{ (\lambda_- - 1)/2 } \right) &  =  &  0
\eea

\vskip.3cm

\noindent
Finally, we get:
\bea
\frac{ \widehat{S}_{(32)}     }{ \widehat{S}_{(01)}      }  & = &
\frac{ \frac{1}{\gamma} + b_1    }  { \frac{1}{\gamma} + a_1  }
\to_{L\to 0} \; \; 1    \label{A23}     \\
\frac{ \widehat{S}'_{(32)}     }{ \widehat{S}'_{(01)}      }  & = &
\frac{  b_1 \psi'_{1 \; (32)} +  b_2 \psi'_{2 \; (32)}   }  { a_1 \phi'_{1 \; (01)} +  a_2 \phi'_{2 \; (01)}  }
\to_{L\to 0}  \;  \; - \;  1  \label{A24}
\eea
When $L\to 0$, we get, for the original graph,  $\widehat{S}$ continuous in $0$ and 
\be\label{A25}
\widehat{S}'_{(01)}      \; + \;
   \widehat{S}'_{(02)}    \;   =   \;  0 
\ee

\vskip1.cm

\section{ Backward Equation: boundary conditions when $p_{01} \ne p_{02}$ }\label{BB}

\vskip.3cm

\begin{figure}
\begin{center}
\includegraphics[scale=.5,angle=0]{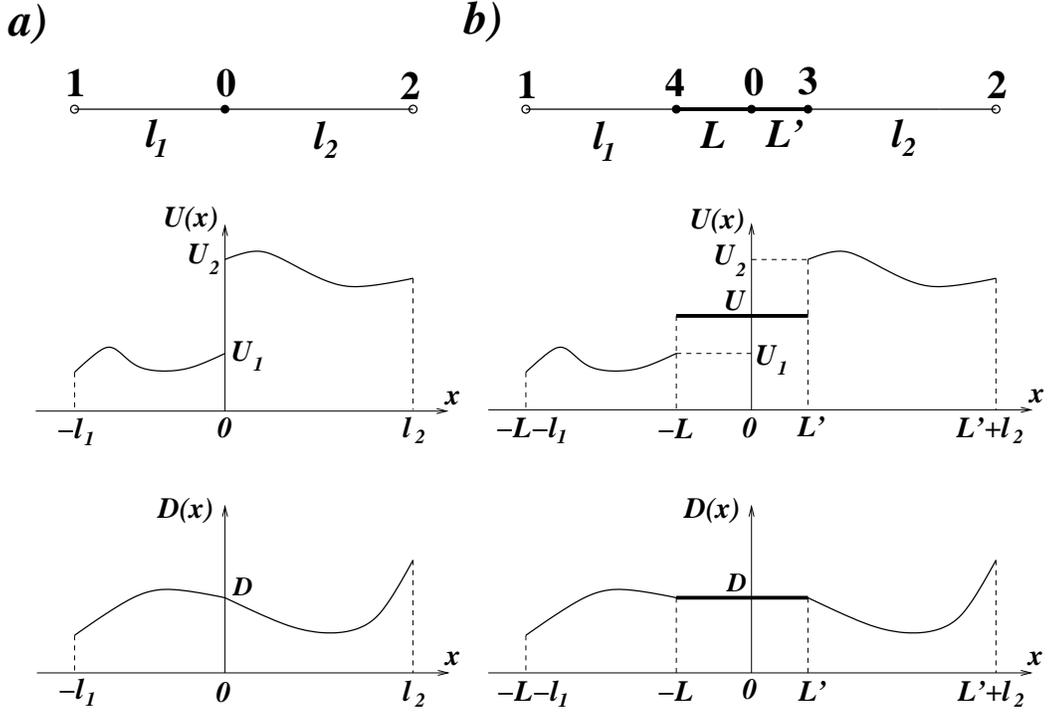}
\end{center}
\caption{ a) The same graph as in Figure \ref{fig6} a) but, this time, $p_{01}\ne
 p_{02} $; in b), $U(x)$ and $D(x)$ are constant 
 on the added links $[40]$ and $[03]$;
 $D(x)$ is still continuous on the modified graph.}\label{fig8} 
\end{figure}

\noindent 
We want to study the case (A) ($U(x)$  discontinuous in $0$) for the
graph of Figure \ref{fig8} a) with, this time, $p_{01}  \ne p_{02}$.

\noindent
Modifying this graph, we obtain a new one, consisting in five
vertices,   
 displayed  in  Figure \ref{fig8} b). Vertices $1$ and $2$ are still
absorbing. We set $ p_{41} = p_{40} =  p_{30} = p_{32} = 1/2  $ 
but  $ p_{04} \ne  p_{03} $ ( $p_{04}= p_{01}$(original graph) and
   $p_{03}= p_{02}$(original graph)). 

\vskip.5cm

\noindent
$U(x)$(resp. $D(x)$) is set equal to some constant $U$(resp. $D$) on
the added links $[40]$ and  $[03]$.
$U(x)$ is discontinuous at vertices $4$ and $3$. 
With the modified graph, we can take advantage of the
computations of the boundary conditions performed at the beginning of
section \ref{bcond} and also in Appendix A.  

\vskip.5cm

\noindent
The solution of equation (\ref{A02}) writes:

\bea
\mbox{on the link } \; [14]  \qquad  \widehat{S}(x_{14}) & = & 1/\gamma + 
   a_1 \phi_1(x_{14})  +  a_2 \phi_2(x_{14})
  \label{B01}   \\    
\mbox{on the link } \; [40]   \qquad \widehat{S}(x_{40}) & = &  1/\gamma + 
b_1  \sinh (\sqrt {\gamma /D} x_{40})       +  b_2 \sinh (\sqrt {\gamma /D} (L-x_{40}))
 \label{B02}   \\  
\mbox{on the link } \; [03] \qquad \widehat{S}(x_{03}) & = &  1/\gamma + 
c_1  \sinh (\sqrt {\gamma /D} x_{03})       +  c_2 \sinh (\sqrt {\gamma /D} (L'-x_{03}))
 \label{B03}   \\  
\mbox{on the link } \; [32]  \qquad  \widehat{S}(x_{32}) & = &   1/\gamma +   
d_1 \psi_1(x_{32})  +  d_2 \psi_2(x_{32})
\label{B04}   
\eea 

\noindent
As before, 
$\phi_{1,2}$ and $\psi_{1,2}$ are solutions of the equation  $(L^+ -
\gamma) \phi = 0$ that satisfy:

\bea
\phi_{1 \; (14)}=0 \quad  \phi_{1 \; (41)}=1 && \phi_{2 \; (14)}=1
\quad  \phi_{2 \; (41)}=0    \label{B05}       \\
\psi_{1 \; (32)}=1 \quad  \psi_{1 \; (23)}=0 && \psi_{2 \; (32)}=0
\quad  \psi_{2 \; (23)}=1  \label{B06}
\eea 

\vskip.5cm   

\noindent
The boundary conditions in $0$ are given  by the beginning of section
\ref{bcbw}. $ \widehat{S}   $ is continuous and 
\be\label{B07}
p_{04} \;  \widehat{S}'_{(04)}  \; + \; p_{03} \;  \widehat{S}'_{(03)}
\; = \; 0
\ee

For vertices $4 $ and  $3  $, we use the result of Appendix A (case
(A)).  $ \widehat{S}   $ is again continuous and 

\bea
 \widehat{S}'_{(41)}   e^{-U_1/D}     \; + \;  \widehat{S}'_{(40)}
     e^{-U/D}  & = & 0 \label{B08} \\
 \widehat{S}'_{(30)}   e^{-U/D}     \; + \;  \widehat{S}'_{(32)}
     e^{-U_2/D}  & = & 0 \label{B09}
\eea
We get the relationships:

\bea
a_2= d_2=-\frac{1}{\gamma   } \qquad 
a_1 =b_2   \sinh (\sqrt {\gamma /D} L)  \qquad d_1 = c_1   \sinh
(\sqrt {\gamma /D} L') &&  \\
b_1   \sinh (\sqrt {\gamma /D} L) =  c_2   \sinh
(\sqrt {\gamma /D} L') &&  \\
p_{04}\left(    -b_1 \cosh (\sqrt {\gamma /D} L) + b_2 \right) \; + \;                           
p_{03}\left(    c_1 -   c_2 \cosh (\sqrt {\gamma /D} L')  \right)      & = &  0  \\
\left(  a_1 \phi'_{1 \; (41)} +  a_2 \phi'_{2 \; (41)}  \right) e^{-U_1/D} \; + \;  
\sqrt {\gamma /D} \left(b_1 - b_2   \cosh (\sqrt {\gamma /D} L)       \right) e^{-U/D} 
& = &  0  \\
\left(  d_1 \psi'_{1 \; (32)} +  d_2 \psi'_{2 \; (32)}  \right) e^{-U_2/D} \; - \;  
\sqrt {\gamma /D} \left(c_1  \cosh (\sqrt {\gamma /D} L')  -c_2 \right) e^{-U/D} 
       & = &  0  
\eea
Solving and taking the limit $L,L' \to 0$, we are left with:

\bea
\frac{\widehat{S}_{(32)}} {\widehat{S}_{(41)}}=\frac{\frac{1}{\gamma} +
   d_1}{\frac{1}{\gamma} + a_1  }  \to_{L,L' \to 0}  &&  1   \label{B10} \\
\frac{\widehat{S}'_{(32)}}{\widehat{S}'_{(41)}}=\frac{
d_1 \psi'_{1 \; (32)} +  d_2 \psi'_{2 \; (32)} }
{  a_1 \phi'_{1 \; (41)} +  a_2 \phi'_{2 \; (41)}}  \to_{L,L' \to 0}
&& e^{ \frac{U_2 - U_1}{D} } \left( \; - \; \frac{p_{04}}{p_{03} \; }
\right)  \label{B11}
\eea
When $L,L' \to 0$, the vertices $3$ and $4$ move to $0$ and we
get\footnote{ Recall that $p_{04}=p_{01}$(original graph)
and $p_{03}=p_{02}$(original graph).}:

\bea
\widehat{S}_{(02)}    & = &      \widehat{S}_{(01)}
\label{B12} \\
p_{02} \;  \widehat{S}'_{(02)} \;  e^{-U_{(02)}/D(0)}     \;  + \;
p_{01} \; \widehat{S}'_{(01)} \; 
e^{-U_{(01)}/D(0)}    & = & 0 \label{B13}
\eea

\vskip.5cm   

\noindent
We also observe that:
\bea
\frac{ \widehat{S}_{(03)} } { \widehat{S}_{(30)}  }  \to_{L,L' \to 0}
&&  1   \label{B100}  \\
\frac{ \widehat{S}'_{(03)} } { \widehat{S}'_{(30)}  }  \to_{L,L' \to 0}
&& - \; 1   \label{B101} 
\eea
Those relationships will show useful in Appendix C.

\vskip1.cm   

\noindent
The solution for the case (B) (Figure 9 a), $p_{01} \ne p_{02}  $) is
immediate.
\begin{figure}
\begin{center}
\includegraphics[scale=.5,angle=0]{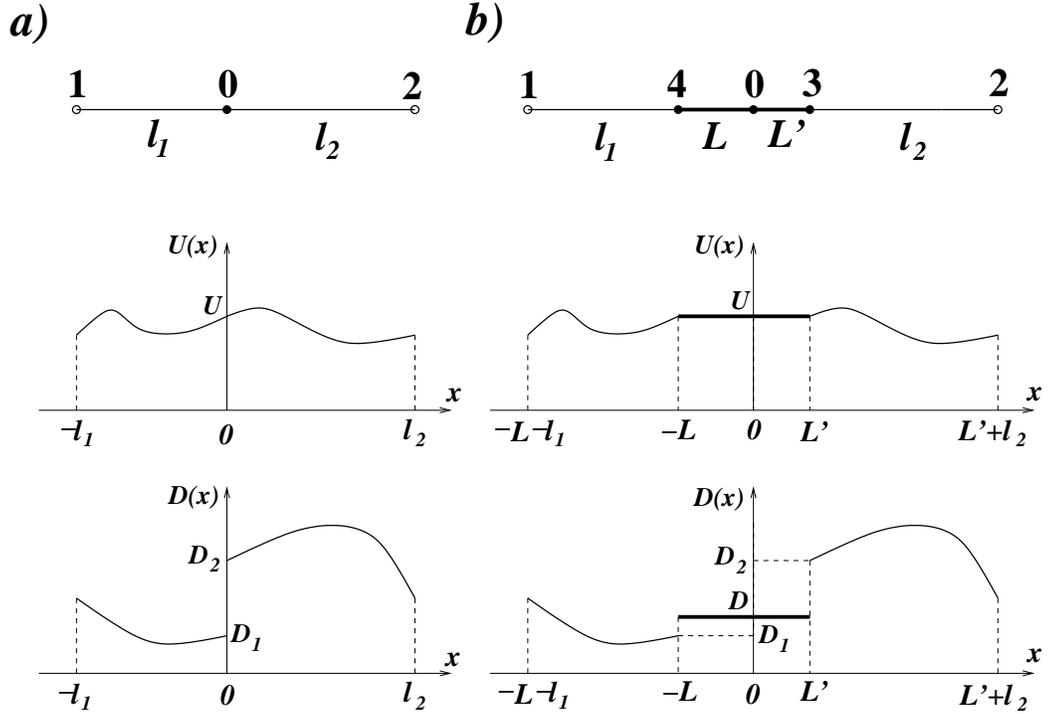}
\end{center}
\caption{ a)  The same graph as in Figure \ref{fig7} a)  but, this time, $p_{01}\ne
 p_{02} $; in b), $U(x)$ and $D(x)$ are constant
on the added links $[40]$ and $[03]$; 
 $U(x)$ is still continuous on the modified graph.  }\label{fig9}
\end{figure}
Indeed, we see that for the modified graph (Figure \ref{fig9}  b)),
 we have, formally, the same solution as equations (\ref{B01})-(\ref{B04})
and the same boundary conditions but, this time, with $U_1=U_2=U$.
 It amounts to drop the exponential factors in (\ref{B08}) and in all
 the equations that follow. Finally, we get that 
 $\widehat{S}$ is continuous and  
\be\label{B20}
p_{02}  \;  \widehat{S}'_{(02)} \;  + \; p_{01} \; \widehat{S}'_{(01)}  = 0 
\ee

\vskip.4cm   

\noindent
Moreover, (\ref{B100}) and (\ref{B101}) still hold.

\vskip1.cm

\section{Backward  Equation:  general  boundary  conditions  }\label{CC}

\vskip.3cm

\begin{figure}
\begin{center}
\includegraphics[scale=.3,angle=0]{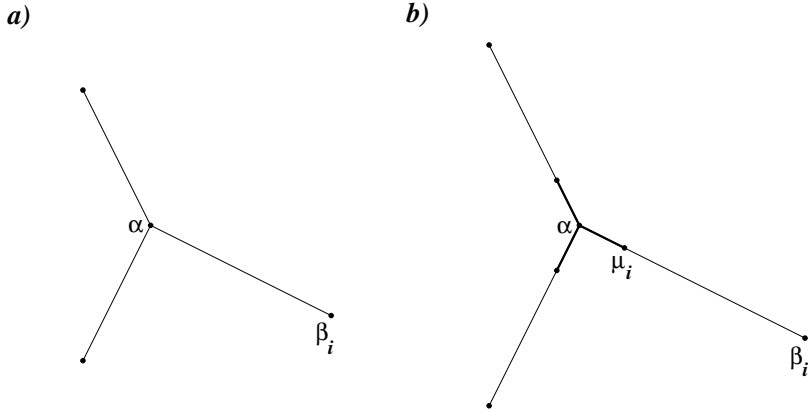}
\end{center}
\caption{a) the vertex $\alpha $ with its nearest-neighbours
  $\beta_i$, $i=1, ...,m_{\alpha }$; b) we have added  the
heavy lines where  the potential is set equal to some constant
$U$ and the diffusion constant is equal to $D(\alpha)$; 
 for the rest of the graph, nothing has been changed. 
For further explanations, see the  text. }\label{bcA}
\end{figure}

\noindent 
For the case (A), let us assume that the $p_{\abi}$'s are not all
equal. In Figure \ref{bcA} a), where a given vertex  $\alpha$ and its
$m_\alpha $  
nearest-neighbours are shown, we suppose that $U(x)$ is discontinuous
in  $\alpha$ ($D(x)$ is continuous, $D(x)\equiv D(\alpha )$ in
$\alpha $)   .

\noindent
In part b), we slightly modify the graph along the same lines as in 
Appendix B.
 We add vertices $\mu_i$ in such a way
that  each  new link $[\mu_i\beta_i ]$ is identical to the original link
  $[\alpha\beta_i ]$ (same length, same potential and diffusion
  constant). Moreover, in the added subgraph (Figure \ref{bcA} b),  heavy
  lines of lengths $L$) the potential and the diffusion constant are
  assumed to be constant (respectively equal to some value $U$ and to
  $D(\alpha)$). So, in the vicinity of $\alpha$, the discontinuities
  of $U(x)$ will occur at the $\mu_i$'s ($D(x)$ will be continuous in
  the same domain). Of course, for the transition probabilities from
  $\alpha $, we choose $p_{\alpha\mu_i} = p_{\abi }  $. By  taking the
  limit $L\to 0$, we will recover the original graph. 

\vskip.6cm

\noindent
Now, for the small subgraph where $U(x) $ and $D(x) $ are constant, we
can take advantage of the result, equation (\ref{back5}), to write:
\be\label{back9}
\sum_{i=1}^{m_{\alpha } } \;  p_{\alpha\mu_i}  \;  P'_{(\alpha\mu_i )} \; = \; 0 
\ee
Moreover, for the vertex $\mu_i$, where
$p_{\mu_i\alpha }=p_{\mu_i\beta_i}=1/2  $, we can use (\ref{back8})
and also (\ref{A14}) (directly established in Appendix A) to get:
\be\label{back10}  
e^{-U/D(\alpha )} P'_{(\mu_i\alpha )} \; + \; e^{-U_{(\mu_i\beta_i )} /D(\alpha )} P'_{(\mu_i\beta_i )}
 \; = \; 0
\ee  

\vskip.2cm

\noindent
Weighting with  $p_{\abi} (\equiv p_{\alpha\mu_i} )$   and summing over $i$, we deduce:
\be\label{back11} 
\sum_{i=1}^{m_{\alpha } } \; p_{\abi}  e^{-U_{(\mu_i\beta_i )} /D(\alpha )} P'_{(\mu_i\beta_i )}
\; = \; - \; e^{-U/D(\alpha )} \left(  
\sum_{i=1}^{m_{\alpha } } p_{\alpha\mu_i} \; P'_{(\mu_i\alpha )}   
\right)
\ee  

\vskip.2cm

\noindent
Now, taking the limit $L\to 0$, we have from (\ref{B101}) (Appendix
B):
\be\label{back12}  
\frac{  P'_{(\mu_i\alpha )}    }    { P'_{(\alpha\mu_i )}   }
\to_{L\to 0} \; - \; 1
\ee
In this limit, the vertex $\mu_i$ moves to $\alpha $ and  we recover the original graph. 
Finally, with (\ref{back9},\ref{back11},\ref{back12}), we obtain, for the
case (A),  the  boundary condition\footnote{Remark that
  (\ref{back13C})
 is unchanged when we add a constant
 to $U(x)$. Now, if we want to consider the case when $U(x)$ and $D(x)$ are
 both  discontinuous at some vertex $\alpha$, we must add, on each
 link, vertices $\mu_i$ and  $\mu'_i$   where either $D(x)$ or $U(x)$ are discontinuous.
 The resulting boundary condition will depend on the repartition of those
 additional vertices. Moreover, inconsistencies will appear when we
 add a constant to  $U(x)$. This is why we say that, in our opinion, this problem is ill-defined.}:
\be\label{back13C} 
\sum_{i=1}^{m_{\alpha } } \; p_{\abi} \;  e^{-U_{(\alpha\beta_i )}
  /D(\alpha )} \; P'_{(\alpha\beta_i )}
 \; = \; 0
\ee

\vskip.2cm

\noindent
Moreover, for the modified graph, $P$ is continuous in $\alpha$ and in
$\mu_i$. From Appendix B, we know that (\ref{B100}):
\be\label{back14}  
\frac{  P_{(\mu_i\alpha )}    }    { P_{(\alpha\mu_i )}   }
\to_{L\to 0}  \; 1
\ee 
That's enough to conclude that, for the original graph,  $P$ is continuous in $\alpha$.

\end{appendix}

\end{document}